\documentclass[prb,twocolumn,superscriptaddress,showpacs]{revtex4-1}
\usepackage{amsmath,graphicx,amssymb,mathrsfs}

\newcount\minute
\newcount\hour
\newcount\hourMins
\def\now
{
  \minute=\time
  \hour=\time \divide \hour by 60
  \hourMins=\hour \multiply\hourMins by 60
  \advance\minute by -\hourMins
  \zeroPadTwo{\the\hour}:\zeroPadTwo{\the\minute}
}
\def\timestamp
{
  \today\ \now
}
\def\today
{
  \zeroPadTwo{\the\day}-\zeroPadTwo{\the\month}-\the\year%
}
\def\zeroPadTwo#1%
{
  \ifnum #1<10 0\fi
  #1%
}
\def \dif {{d}}
\def \Dif {\mathcal{D}}

\begin{document}

\title{Superfluid/Bose-glass transition in one dimension}

\author{Zoran Ristivojevic}
\affiliation{Centre de Physique Th\'{e}orique, Ecole Polytechnique, CNRS, 91128 Palaiseau, France }
\author{Aleksandra Petkovi\'{c}}
\affiliation{Laboratoire de Physique Th\'{e}orique, IRSAMC, CNRS and Universit\'{e} de Toulouse, UPS, 31062 Toulouse,
France}
\author{Pierre Le Doussal}
\affiliation{Laboratoire de Physique Th\'{e}orique--CNRS, Ecole Normale Sup\'{e}rieure, 24 rue Lhomond, 75005 Paris, France}
 \author{Thierry Giamarchi}
\affiliation{DPMC-MaNEP, University of Geneva, 24 Quai Ernest-Ansermet, CH-1211 Geneva, Switzerland}

\begin{abstract}
We consider a one-dimensional system of interacting bosons in a random potential. At zero temperature, it can be either in the superfluid or in the insulating phase. We study the transition at weak disorder and moderate interaction. Using a systematic approach, we derive the renormalization group equations at two-loop order and discuss the phase diagram. We find the universal form of the correlation functions at the transitions and compute the logarithmic corrections to the main universal power-law behavior. In order to mimic large density fluctuations on a single site, we study a simplified model of disordered two-leg bosonic ladders with correlated disorder across the rung. Contrarily to the single-chain case, the latter system exhibits a transition between a superfluid and a localized phase where the exponents of the correlation functions at the transition do not take universal values.
\end{abstract}

\date{\timestamp}


\maketitle

\section{Introduction}

The combined effect of disorder and interactions is one of the most fascinating problems of quantum correlated systems. Indeed, interactions can lead to collective effects such as superconductivity or superfluidity for which the many-body function is known to be resistant to disorder. On the other hand, for single particles the quantum nature of the problem is known to strongly reinforce the effects of disorder leading to the celebrated Anderson localization, \cite{Anderson58PhysRev}, for which the system is an insulator. One can thus expect a fierce competition between these two phenomena.

One of  the systems for which this competition manifests in its strongest possible way are disordered bosons. Indeed, in this case one can expect a competition between superfluidity and Anderson localization. In one dimension it was shown by
a renormalization group analysis \cite{Giamarchi+88} that disordered interacting bosons would undergo a phase transition between a superfluid and a localized phase. This transition and this phase, nicknamed Bose glass, was also shown \cite{Fisher+89} by scaling arguments to exist in higher dimensions. The phase diagram of disordered interacting bosons in one dimension possesses several remarkable features. First, the interactions lead to a reentrant superfluid phase, leading first to a delocalization from Bose glass to superfluid, when increased, then to a second transition at stronger interactions between the superfluid and the Bose glass.  \cite{Giamarchi+88} Second, the transition at moderate interactions is in the Berzinskii-Kosterlitz-Thouless universality class, and {universal} exponents exists for the various correlation functions, in particular the single-particle one, at the transition. Further analyses of the transition in terms of vortex proliferation \cite{svistunov+96PhysRevB.53.13091} confirmed and generalized this result and numerical studies of disordered bosons \cite{Prokofev+98PhysRevLett.80.4355,Rapsch+99,Pollet+09} confirmed both the reentrant nature of the phase diagram and the nature of the moderate interaction transition.\cite{refael+13,pollet2013review} Extensions of this transition to finite temperature have also recently attracted a great deal of attention. \cite{aleiner2010finite}

On the experimental front, the superfluid/Bose-glass transition has regained a considerable interest thanks to cold atomic gases which have provided remarkable realization of this problem \cite{Billy+08,Roati+08,Pasienski+10} as well as realization of disordered bosons by magnetic insulators. \cite{giamarchi+08naturephysics,Zheludev+10PhysRevB.81.060410,Yamada+11PhysRevB.83.020409,Zapf+12nature,zheludev2013dirty,ward2013spin}
This has led to a hunt for the Bose glass, as well as new questions on the phase diagram. In particular new studies focused on the low interaction, strong disorder case. \cite{Lugan+11,Fontalesi+11PhysRevA.83.033626} In this regime, a real-space renormalization group study by \citet{Altman+10} of a related disordered Josephson junction array model found again a Berzinskii-Kosterlitz-Thouless transition, but with disorder-dependent exponents at the transition. These results which are still being debated \cite{Svistunov05,Vojta,Pollet13,Pielawa+13,pollet+14PhysRevB.89.054204} nevertheless strongly suggest the existence of a different universality class for the weak interaction Bose-glass/superfluid transition than for the moderate interaction one, with universal exponents, and thus a richer phase diagram, a possibility already discussed in Ref.~\onlinecite{Giamarchi+88}.

In order to analyze further the moderate interaction side of the phase diagram we had in a recent paper \cite{PRLours} analyzed the moderate interaction superfluid/Bose-glass transition at the next order in the renormalization group analysis. This study confirmed the universality of the exponent at the transition, in a way consistent with the previous studies, and that non-universal terms such as those appearing in slightly different disordered problems such as the Cardy-Ostlund model \cite{Cardy+82} were indeed not generated. It also allowed a precise calculation of the correlation functions at the transition including their logarithmic corrections to the main power law.

In this paper we give a detailed account of the renormalization technique used in Ref.~\onlinecite{PRLours}, since it can be useful to related problems. We also detail the calculation of the correlation functions. In addition, we generalize the study to more complex models than a single chain of bosons. In particular, we investigate the case of a disordered bosonic ladder with correlated disorder along the rung (for the uncorrelated case see Ref.~\onlinecite{orignac+giamarchi}). This system caricatures a bosonic system where density fluctuations on a single rung can be larger than for a single chain. We show that in that case the exponent is not universal at the transition in contrast to the case of the single bosonic chain.

The plan of the paper is as follows. In Sec.~\ref{sec:model} we introduce the model that we treat using the replica method. We calculate the correlation functions with respect to the harmonic part of the model. In Sec.~\ref{sec:effectiveaction} we calculate the effective action at weak disorder. In Sec.~\ref{sec:RG} we obtain the renormalization group flow equations, and solve them finding the phase diagram. In Sec.~\ref{sec:generator} we calculate the generator of connected correlations that enables us to calculate the density-density correlation function in Sec.~\ref{sec:correlations} and the single-particle correlation function  in Sec.~\ref{sec:singleparticleCF}. In Sec.~\ref{sec:ladder} we consider a two-leg ladder problem of disordered bosons, which exhibits nonuniversal exponents at the superfluid/Bose-glass transition. Section~\ref{sec:discussion} contains the discussions, which are followed by conclusions. Some details of calculations are presented in Appendices \ref{appendixa1}, \ref{appendixb1}, \ref{appendixb2}, and \ref{appendixladder}.

\section{Model\label{sec:model}}

We study the system of one-dimensional interacting bosons in a disordered potential. At low energies, the clean system can be described by the Tomonaga-Luttinger Hamiltonian \cite{Haldane81,Giamarchi,cazalilla+2011RMP}
\begin{align}\label{eq:llham}
H_0=\frac{\hbar}{2\pi}\int\dif x\left\{vK[\partial_x{\theta}(x)]^2 +\frac{v}{K}[\partial_x{\varphi}(x)]^2 \right\}.
\end{align}
Here $v$ denotes the sound velocity and $K$ is a dimensionless Luttinger liquid parameter that accounts for the interaction strength. The fields $\varphi$ and $\theta$ satisfy the bosonic commutation relation $[{\varphi}(x),\partial_y{\theta}(y)]=i\pi\delta(x-y)$.

The parameters $v$ and $K$ of the phenomenological Hamiltonian (\ref{eq:llham}) can be related to the parameters of a specific microscopic Hamiltonian. For example, the microscopic Lieb-Liniger model \cite{lieb1963exact} describes bosons with local repulsion of the strength $g$. At low energies, the latter model can be represented by the Hamiltonian (\ref{eq:llham}). The parameter $K$ of Eq.~(\ref{eq:llham}) is such that at very strong interaction when $g\to\infty$, we have $K\to 1$,\cite{ristivojevic14} while in the opposite case of weak interaction $g\to 0$, we have  $K\to\infty$.\cite{cazalilla+2011RMP} Therefore, $K\ge 1$ at any strength of the repulsive contact interaction. The parameter regime $K<1$ can be reached only by including longer range interaction in the microscopic model. In Galilean invariant microscopic models, which include the Lieb-Liniger one, the product $vK$ is fixed and determined by ratio between the mean density and the mass of physical particles forming the bosonic system.

The density of bosonic particles can be expressed in terms of the displacement fields $\varphi$ as \cite{haldane81prl}
\begin{align}\label{density}
\rho(x)=\rho_0-\frac{1}{\pi}\partial_x\varphi(x) +2\rho_2\cos[2\varphi(x)-2\pi\rho_0x].
\end{align}
In Eq.~(\ref{density}), $\rho_0$ denotes the mean density while the remaining terms account for the density fluctuations. The second term in Eq.~(\ref{density}) describes long-wavelength fluctuations, while the oscillatory term describes the density fluctuations around the wave vector $\pm 2\pi\rho_0$. The constant $\rho_2$ is nonuniversal and depends on microscopic details. \cite{cazalilla+2011RMP}

In the following, we consider the system of disordered bosons and account for the effects of the disorder via its coupling to the particle density (\ref{density}). Thus, we include additional terms in the model that are of the form
\begin{align}\label{eq:Hd}
H_{d}= \int\dif x\left\{-\frac{1}{\pi}\eta(x)\partial_x\varphi+\rho_2\left[\xi^*(x) {e}^{i2\varphi}+\text{h.c.}\right]\right\}.
\end{align}
Here we distinguish the so-called forward $\eta$ and backward $\xi$ scattering caused by the disorder potential. They respectively have Fourier components around the wavevectors  $0$ and $\pm 2\pi\rho_0$.\cite{Giamarchi} We consider Gaussian disorder where the fields $\eta$ and $\xi$ have mean values zero and correlations
\begin{gather}\label{definitionDf}
\overline{\eta(x)\eta(x')}=\hbar^2D_f\delta(x-x'),\\ \label{definitionD} \overline{\xi(x)\xi^*(x')}=\hbar^2D_b\delta(x-x'),\\\label{eq:mixed}
\overline{\eta(x)\xi(x')}=\overline{\xi(x)\xi(x')}=0.
\end{gather}
We therefore assume that $D_f$ and $D_b$ characterize the disorder strength. Here and in the following $\overline{.\phantom{1}.\phantom{1}.}$ denotes the disorder average. The total Hamiltonian is
\begin{align}\label{Hamiltonian}
H=H_0+H_{d}.
\end{align}

In order to treat the effects of disorder we employ the replica method.\cite{Giamarchi} We reexpress the free energy $F=-T \ln{Z}$, where $Z$ denotes the partition function, as
$F=-T\lim_{n\to 0}(Z^n-1)/n$. The latter form enables us to introduce the replicated action  $S_{\rm rep}$ that contains the disorder average. It is defined via the relation
\begin{align}
\overline{Z^n}=\int \left(\prod_{\alpha=1}^n\mathcal{D}\theta_{\alpha}\mathcal{D}\varphi_{\alpha} \right) e^{-S_{\mathrm{rep}}/\hbar},
\end{align}
where $\varphi_\alpha$ and $\theta_{\alpha}$ are the replicated fields. A technical difficulty of the replica method is that the replicated action depends on $2n$ fields, in contrast to two fields present in the initial Hamiltonian.

For the model (\ref{Hamiltonian}), we conveniently split the replicated action as
\begin{align}\label{Srep}
S_{\mathrm{rep}}=S_0^{}+S_f^{}+S_b^{}.
\end{align}
The first term in Eq.~(\ref{Srep}) is quadratic as it arises from $H_0$. It has the form
\begin{align}\label{S0-rep-full}
\frac{S_0^{}}{\hbar}=&\frac{1}{2\pi}\sum_{\alpha} \int
\dif x\dif \tau\bigg[vK(\partial_x\theta_\alpha)^2 +\frac{v}{K}(\partial_x\varphi_{\alpha})^2\notag\\ &+2i(\partial_x\theta_\alpha)(\partial_\tau\varphi_\alpha)\bigg].
\end{align}
The second term in Eq.~(\ref{Srep}) originates from $\eta$-dependent part of the  disordered potential (\ref{eq:Hd}). It is also quadratic, but off-diagonal in replica space,
\begin{align}\label{Sf-rep}
\frac{S_f^{}}{\hbar}=-\frac{D_f}{2\pi^2} \sum_{\alpha \beta}\int
\dif x\dif\tau\dif\tau' [\partial_x\varphi_\alpha(x,\tau)][\partial_x\varphi_\beta(x,\tau')].
\end{align}
The third term in Eq.~(\ref{Srep}) is anharmonic and originates from the disordered potential expressed through the $\xi$ field in Eq.~(\ref{eq:Hd}). It describes the effects of backward scattering on the disordered potential and reads
\begin{align}\label{Sb-rep}
\frac{S_b^{}}{\hbar}=-\rho_2^2 D_b \sum_{\alpha\beta}\int\dif x\dif\tau\dif\tau' \cos[2\varphi_\alpha(x,\tau)-2\varphi_\beta(x,\tau')].
\end{align}
We therefore see that Gaussian disorder enters into the replicated action through the parameters describing its variance, $D_f$ and $D_b$, see Eqs.~(\ref{definitionDf}) and (\ref{definitionD}).

The disorder potential does not couple to the phase field $\theta$. It is therefore sometimes useful to integrate out the fields $\theta_\alpha$ from the replicated action. As it involves Gaussian integrations, the resulting quadratic part in the action can be found exactly,
\begin{align}\label{S0-rep}
\frac{S_0^{}}{\hbar}=&\frac{v}{2\pi K}\sum_{\alpha} \int
\dif x\dif \tau\bigg[(\partial_x\varphi_{\alpha})^2+ \frac{1}{v^2}(\partial_\tau\varphi_\alpha)^2 +m^2\varphi_\alpha^2\bigg].
\end{align}
In Eq.~(\ref{S0-rep}) we have introduced a small mass $m$ that represents an infrared cutoff. On some occasions, we perform  calculations at finite $m$, taking the limit $m\to 0$ at the end. An example is the evaluation of the effective action, see Sec.~\ref{sec:effectiveaction}. However, for calculation of some quantities we can use $m=0$ from the beginning. We encounter this situation when calculating the density-density correlation function and the single-particle correlation function. We emphasize that in the most part of this article we use the expression (\ref{S0-rep}) rather than (\ref{S0-rep-full}). The latter expression we only employ to calculate the single-particle correlation function. This is expected, as $\theta$ fields are involved in its definition.

In this work we study the case of small anharmonic term (\ref{Sb-rep}) that enables us to use perturbation theory. We therefore need the correlation function
\begin{align}\label{Gabdef}
G_{\alpha\beta}(x,\tau)= \left\langle\varphi_{\alpha}(x,\tau)\varphi_{\beta}(0,0) \right\rangle,
\end{align}
where $\langle\ldots\rangle$ denotes an average with respect to the quadratic action $S_0^{}+S_f^{}$. Introducing the Fourier transform as
\begin{align}
\varphi_\alpha(x,\tau)=\int\frac{\dif k\dif\omega}{(2\pi)^2} {e}^{i(k x+\omega\tau)}\varphi_\alpha(k,\omega),
\end{align}
we easily diagonalize Eqs.~(\ref{S0-rep}) and (\ref{Sf-rep}), and obtain
\begin{align}\label{S0SF}
&\frac{S_0^{}+S_f^{}}{\hbar}=\frac{1}{2}\sum_{\alpha\beta}\int\frac{\dif k\dif\omega}{(2\pi)^2}\frac{\varphi_\alpha(k,\omega) \varphi_\beta(-k,-\omega)}{G_{\alpha\beta}(k,\omega)}.
\end{align}
The propagator $G_{\alpha\beta}(k,\omega)=\langle\varphi_\alpha(k,\omega) \varphi_\beta(-k,-\omega)\rangle$ in the previous equation is given by \footnote{We used the inversion formula for the matrix $M_{\alpha\beta}=A\delta_{\alpha\beta}+B$ which reads $(M_{\alpha\beta})^{-1}=\frac{1}{A}\delta_{\alpha\beta}+ \frac{1}{n}\left(\frac{1}{A+nB}-\frac{1}{A}\right)= \frac{1}{A}\delta_{\alpha\beta}-\frac{B}{A^2}+\mathcal{O}(n)$.
}
\begin{align}\label{Galphabeta}
G_{\alpha\beta}(k,\omega)=&\left[\frac{\delta_{\alpha\beta}v}{\pi K}\left(k^2+\frac{\omega^2}{v^2}+m^2\right)-\frac{2D_f}{\pi} k^2\delta(\omega)\right]^{-1}\notag\\
=&\frac{\pi K/v}{k^2+\frac{\omega^2}{v^2}+ m^2}\delta_{\alpha\beta}+\frac{2\pi K^2D_f}{v^2} \notag\\
&\times \frac{k^2\delta(\omega)} {\left(k^2+\frac{\omega^2}{v^2}+m^2\right)^2}+\mathcal{O}(n).
\end{align}
After performing inverse Fourier transform and setting the number of replicas $n$ to zero, for the correlation function (\ref{Gabdef}) we obtain
\begin{align}\label{Gab}
G_{\alpha\beta}(x,\tau)= \delta_{\alpha\beta}G(x,\tau)+G_0(x),
\end{align}
where
\begin{gather}\label{G}
G(x,\tau)=\frac{K}{2}K_0\left(m\sqrt{x^2+v^2\tau^2+a^2}\right),\\
\label{G0def}
G_0(x)=\frac{K^2D_f}{4v^2}\frac{{e}^{-m|x|}}{m}(1-m|x|).
\end{gather}
In Eq.~(\ref{G}) we have introduced the parameter $a$ as an ultraviolet cutoff, while $K_0$ denotes the modified Bessel function of the second kind.\cite{Amit+80} In the limit of small distances $\sqrt{x^2+v^2\tau^2}\ll (cm)^{-1}$, the correlation functions (\ref{G}) and (\ref{G0def}) become
\begin{gather}\label{Gsmall}
G(x,\tau)=-\frac{3+2\delta}{8}\ln \left[c^2 m^2(x^2+v^2\tau^2+a^2)\right],\\
\label{G0small}
G_0(0)-G_0(x)=\frac{K^2 D_f}{2v^2}|x|,\quad |x|\ll m^{-1}.
\end{gather}
Here $c=e^{\gamma_E}/2$ is the constant, where $\gamma_E$ denotes the Euler constant. In Eq.~(\ref{Gsmall}) we have conveniently introduced
\begin{align}\label{delta definition}
\delta=K-3/2.
\end{align}
The parameter $\delta$ of Eq.~(\ref{delta definition}) is introduced as its renormalized value measures the distance from the phase transition between the superfluid and the disordered phase that occurs at moderate interaction corresponding\cite{Giamarchi+88} to the renormalized Luttinger liquid parameter $3/2$. In the superfluid phase, the renormalized parameter corresponding to $\delta$ is positive. The strength of the anharmonic disordered potential (\ref{Sb-rep}) and $\delta$ are small parameters in our study.

\section{Effective action\label{sec:effectiveaction}}

In this section we derive the renormalization group scaling equations at two-loop order. We use the standard field-theoretical method, which is particularly suitable for calculations beyond lowest order.\cite{Zinn-Justin,Amit,Amit+80} While being somewhat different from the standard Kadanoff-Wilson approach popular in condensed matter physics, it yields equivalent results.\cite{dombgreen-vol6,jona} The field-theoretical technique is used in a number of earlier studies. To mention the most relevant to our study, Ref.~\onlinecite{Amit+80} studied the sine-Gordon model at two-loop order, while Ref.~\onlinecite{ristivojevic+12PhysRevB.86.054201} considered the random-phase sine-Gordon model at two-loop order. Here we study the quantum version of the random-phase sine-Gordon model.

The main quantity needed to derive the scaling equations is the effective action. For an action $S(\varphi)$, it is defined as $\Gamma(\varphi)=J\varphi-W(J)$, where $W(J)$ is the generator of connected correlations defined as
$
{e}^{{W}(J)}=\int\Dif\varphi {e}^{-S(\varphi)/\hbar+J\varphi}.
$
Using $J(x)=\frac{\delta \Gamma}{\delta \varphi(x)}$,
we obtain an implicit equation for the effective action \cite{Zinn-Justin}
\begin{align}\label{Gamma}
{e}^{-\Gamma(\varphi)}=\int\Dif\chi {e}^{-\frac{S(\varphi+\chi)}{\hbar}+\int\dif x\chi(x)\frac{\delta \Gamma}{\delta \varphi(x)}}.
\end{align}
Equation (\ref{Gamma}) applies for an arbitrary action $S$.

We consider the model (\ref{Srep}) at weak disorder that enables us to solve the implicit equation (\ref{Gamma}) using a perturbation theory. It is controlled by the small parameter $D_b$ that determines the strength of the anharmonic term (\ref{Sb-rep}). Up to an additive constant, the effective action for the model (\ref{Srep}) is given by the expression\cite{ristivojevic+12PhysRevB.86.054201}
\begin{align}\label{Gamma pert}
\Gamma=\frac{S_0+S_f}{\hbar} +\Gamma_1+\Gamma_2+\mathcal{O}(D_b^3),
\end{align}
where
\begin{align}
\label{Gamma1def}
\Gamma_1=&\frac{1}{\hbar}\langle S_b(\varphi+\chi)\rangle_\chi,\\
\label{Gamma2def}
\Gamma_2=&-\frac{\langle S_b^2(\varphi+\chi)\rangle_\chi}{2\hbar^2}+\frac{1}{2}\Gamma_1^2 \notag\\ &+\frac{1}{2}\sum_{\alpha\beta}\int\dif x\dif\tau\dif x'\dif\tau' G_{\alpha\beta}(x-x',\tau-\tau') \notag\\ &\times\frac{\delta\Gamma_1}{\delta\varphi_\alpha(x,\tau)} \frac{\delta\Gamma_1}{\delta\varphi_\beta(x',\tau')}.
\end{align}
Here $\langle\cdots\rangle_\chi$ denotes an average with respect to the quadratic action $S_0(\chi)+S_f(\chi)$. For example, the average of the functional $F(\varphi+\chi)$ is defined as
\begin{align}
\langle F(\varphi+\chi)\rangle_\chi=\frac{\int\Dif \chi F(\varphi+\chi) e^{-\frac{S_0(\chi)+S_f(\chi)}{\hbar}} }{\int\Dif \chi e^{-\frac{S_0(\chi)+S_f(\chi)}{\hbar}}}.
\end{align}

Now we evaluate the effective action. Using Eq.~(\ref{Gamma1def}), we easily obtain
\begin{align}\label{Gamma1}
\Gamma_1=&-\mathscr{B}\sum_{\alpha\beta} \int\dif x\dif\tau\dif\tau' \bigg\{\left[{e}^{4G(0,\tau-\tau')}-1\right] \delta_{\alpha\beta}+1\bigg\}\notag\\
&\times \cos[2\varphi_\alpha(x,\tau)-2\varphi_\beta(x,\tau')],
\end{align}
where $\mathscr{B}=\rho_2^2D_b{e}^{-4G(0,0)}$. We should note that from the correlation function (\ref{Gab}) only the diagonal part in replica indices (\ref{G}) enters the result (\ref{Gamma1}). This is because the off-diagonal part (\ref{G0def}) does not depend on the imaginary time. As will become obvious below in this and in the next section, the lowest order term $\Gamma_1$ contains all the information necessary to obtain the scaling equations of Giamarchi and Schulz. \cite{Giamarchi+87EPL}

The second order term in the expansion of the effective action is given by Eq.~(\ref{Gamma2def}). After a careful calculation, we express the final result as a sum $\Gamma_2=\sum_{j=1}^3\Gamma_2^{(j)}$, where $j$ denotes the number of different replica indices contained in  $\Gamma_2^{(j)}$, which read
\begin{widetext}
\begin{align}
\label{Gamma21}
\Gamma_2^{(1)}=&-\frac{1}{2}\mathscr{B}^2\sum_\alpha \int_{x,\tau,\tau',\atop x_1,\tau_1,\tau_1'} f_1(x-x_1,\tau,\tau',\tau_1,\tau_1') \cos[2\varphi_\alpha(x,\tau)-2\varphi_\alpha(x,\tau')+ 2\varphi_\alpha(x_1,\tau_1)-2\varphi_\alpha(x_1,\tau_1')],\\
\label{Gamma22}
\Gamma_2^{(2)}=&-\mathscr{B}^2\sum_{\alpha\beta} \sum_{s=\pm1}\int_{x,\tau,\tau',\atop x_1,\tau_1,\tau_1'} \bigg\{f_2(x-x_1,\tau-\tau_1',\tau-\tau_1) \cos[2\varphi_\alpha(x,\tau)+2\varphi_\alpha(x_1,\tau_1) -2\varphi_\alpha(x_1,\tau_1') -2\varphi_\beta(x,\tau')]\notag\\
&\qquad\qquad+\frac{1}{2}f_3(x-x_1,\tau-\tau_1,\tau'-\tau_1',s) \cos[2\varphi_\alpha(x,\tau)-2s\varphi_\alpha(x_1,\tau_1)-2\varphi_\beta(x,\tau') +2s\varphi_\beta(x_1,\tau_1')]\bigg\},\\
\label{Gamma23}
\Gamma_2^{(3)}=&-
\mathscr{B}^2\sum_{{\alpha\beta\gamma}} \sum_{s=\pm1}\int_{x,\tau,\tau',\atop x_1,\tau_1,\tau_1'} f_4(x-x_1,\tau-\tau_1,s) \cos[2\varphi_\alpha(x,\tau') -2\varphi_\beta(x,\tau)+2s\varphi_\beta(x_1,\tau_1) -2s\varphi_\gamma(x_1,\tau_1')].
\end{align}
\end{widetext}
Here we have used the shorthand notation $\int_{x,\tau,\ldots}f(x,\tau,\ldots)$ to denote $\int\dif x\dif\tau\ldots f(x,\tau,\ldots)$. The functions $f_2$, $f_3$, and $f_4$ in the previous equations are
\begin{align}\label{f2fin}
f_2(x,\tau,\tau')=&{e}^{4G(x,\tau)-4G(x,\tau')+4G(0,\tau-\tau')} -{e}^{4G(x,\tau)}\notag\\ &-{e}^{-4G(x,\tau')}
-{e}^{4G(0,\tau-\tau')} +2\notag\\
&+4[{e}^{4G(0,\tau-\tau')}-1][G(x,\tau')-G(x,\tau)],\\
f_3(x,\tau,\tau',s)=&[{e}^{4sG(x,\tau)}-1][{e}^{4sG(x,\tau')}-1],\\ f_4(x,\tau,s)=&{e}^{4sG(x,\tau)}-4sG(x,\tau)-1.
\end{align}
The function $f_1$ in $\Gamma_2^{(1)}$ [see Eq.~(\ref{Gamma21})] is cumbersome and not necessary for the present study. Namely, the term $\Gamma_2^{(1)}$ leads to the correction proportional to  $D_b^2$ in the scaling equation for $\delta$ [see the last term $\mathcal{O}(\mathscr{D}_R^2)$ in Eqs.~(\ref{delta})]. This correction is important at three-loop order calculation and thus $f_1$ is not needed here.

The effective action of the model $S_0+S_f+S_b$ [Eqs.~(\ref{S0-rep}), (\ref{Sf-rep}), and (\ref{Sb-rep})] up to second order in the anharmonic coupling $D_b$ is given by Eq.~(\ref{Gamma pert}), where $\Gamma_1+\Gamma_2$ is the sum of the terms in Eqs.~(\ref{Gamma1})-(\ref{Gamma23}).
It contains all the information about critical properties of our model at two-loop order. In order to derive the scaling equations one should find the most relevant operators in the effective action. In the limit when the ultraviolet cutoff $a$ goes to zero, those operators contain divergent prefactors.

The most relevant contributions from Eq.~(\ref{Gamma1}) are
\begin{align}\label{Gamma1 fin}
{\Gamma}_1=&2\mathscr{B}a_1\sum_\alpha \int\dif x\dif\tau [\partial_\tau\varphi_\alpha(x,\tau)]^2 \notag\\&-\mathscr{B}\sum_{\alpha\beta} \int_{x,\tau,\tau'}\cos[2\varphi_\alpha(x,\tau)-2\varphi_\beta(x,\tau')]+
\ldots
\end{align}
where
\begin{align}\label{a1def}
a_1=&\int \dif\tau \tau^2\left[{e}^{4G(0,\tau)}-1\right],
\end{align}
while the ellipsis denotes many irrelevant operators.

The most relevant terms from $\Gamma_2$ are
\begin{align}\label{eq:divGamma2}
 {\Gamma}_2=&-2b_2\mathscr{B}^2
\sum_{\alpha\beta}\int\dif x\dif\tau\dif\tau'[\partial_x\varphi_\alpha(x,\tau)] [\partial_x\varphi_\beta(x,\tau')]\notag\\ &-2b_1\mathscr{B}^2\sum_{\alpha\beta}\int\dif x\dif\tau\dif\tau' \cos[2\varphi_\alpha(x,\tau)-2\varphi_\beta(x,\tau')]\notag\\ &+\mathcal{O}(\mathscr{B}^2) \sum_\alpha \int\dif x\dif\tau [\partial_\tau\varphi_\alpha(x,\tau)]^2+\ldots,
\end{align}
As will be discussed below, the last term Eq.~(\ref{eq:divGamma2}) corresponds to a higher order contribution in the scaling equation for $\delta$. The coefficients $b_1$ and $b_2$ in Eq.~(\ref{eq:divGamma2}) read
\begin{gather}\label{defb1}
b_1=\int\dif x\dif\tau\dif\tau'f_2(x,\tau,\tau+\tau'),\\
\label{defb2}
b_2=\int\dif x\dif\tau\dif\tau'x^2f_3(x,\tau,\tau',1).
\end{gather}
Considering $\Gamma_2^{(2)}$ and performing the derivative expansion, one would naively expect to find  the operator with two replica indices of the form
\begin{align}\label{newoperator}
\sum_{\alpha\beta}\int\dif x\dif x'\dif\tau[\partial_\tau\varphi_\alpha(x,\tau)] [\partial_\tau\varphi_\beta(x',\tau)].
\end{align}
However, the latter operator is not generated in the expanded effective action as a consequence of the equality $f_3(x,\tau,\tau',1)=f_3(x,-\tau,-\tau',1)$. The absence of the operator (\ref{newoperator}) has important consequences for the universality of the exponent of the singe particle correlation function, as we discuss in the following.

The three-replica part (\ref{Gamma23}) contains two contributions that turn out to be unimportant. Namely, for $s=1$ after a gradient expansion a free sum over $\beta$ index delivers a factor $n$, that vanishes in the replica limit. For $s=-1$, one finds the operator of the form $\cos(2\varphi_\alpha+2\varphi_\gamma-4\varphi_\beta)$ where for simplicity we omitted the arguments in replica fields. Such term is irrelevant close to the critical point and has a nondivergent prefactor when $a\to 0$.

Collecting the most relevant terms given by Eqs.~(\ref{Gamma1 fin}) and (\ref{eq:divGamma2}), the effective action (\ref{Gamma pert}) becomes
\begin{widetext}
\begin{align}\label{Gammafinal}
\Gamma=&\sum_{\alpha} \int\dif x\dif\tau\left\{\frac{v}{2\pi K}\left[(\partial_x\varphi_\alpha)^2+m^2(\varphi_\alpha)^2\right]+ \left[\frac{1}{2\pi  K v }+2\mathscr{B}a_1+\mathcal{O}(\mathscr{B}^2)\right] (\partial_\tau\varphi_\alpha)^2\right\} -\left[\frac{D_f}{4\pi^2}+2\mathscr{B}^2b_2\right]\notag\\
&\times\sum_{\alpha\beta}\int\dif x\dif\tau\dif\tau'[\partial_x\varphi_\alpha(x,\tau)] [\partial_x\varphi_\beta(x,\tau')]
-\left(\mathscr{B}+2b_1\mathscr{B}^2\right)\sum_{\alpha\beta}\int\dif x\dif\tau\dif\tau' \cos[2\varphi_\alpha(x,\tau)-2\varphi_\beta(x,\tau')].
\end{align}
\end{widetext}

The parameters $a_1$, $b_1$, and $b_2$ in Eq.~(\ref{Gammafinal}) can be evaluated in a series expansion in the small parameter $\delta$ that is defined in Eq.~(\ref{delta definition}). The details of calculation are presented in Appendices \ref{appendixa1},\ref{appendixb1}, and \ref{appendixb2}, while here we state the final results. In the limit $a\to 0$ we find
\begin{align}\label{a1}
a_1=\frac{-2\lambda+ \delta\lambda^2-4(\ln2-1)\delta\lambda+\mathcal{O}(\delta^2) +2c_1}{2(cmv)^3},
\end{align}
where $\lambda=\ln (c^2m^2a^2)$ and $c$ is the constant defined earlier. In Eq.~(\ref{a1}), $c_1$ is a constant. The other two terms are
\begin{gather}\label{b1}
b_1 =\frac{\pi\left[9\lambda^2+2(51-54\ln2)\lambda +\mathcal{O}(\delta)+4c_2\right]}{8v^2(cm)^3},\\
\label{b2}
b_2 =\frac{2\pi}{v^2(cm)^6}\frac{1}{a} \left[1+\mathcal{O}(\delta)\right]+\frac{c_3}{v^2m^5}. \end{gather}
The constants $c_1$, $c_2$, and $c_3$ are nonuniversal since they depend the choice of the infrared cutoff function in Eq.~(\ref{S0-rep}). We will see below that only $c_1$ appears in the renormalization group equations. In general, the appearance of nonuniversal terms in renormalization group equations is not surprising, since other low-dimensional models also contain them. Some examples are the sine-Gordon model \cite{Amit+80} and the Cardy-Ostlund model.\cite{ristivojevic+12PhysRevB.86.054201} However, those nonuniversal terms do not appear in physical observables  at the transition. In the present case an example is the single particle correlation function that does not depend on $c_1$, as shown in Sec.~\ref{sec:singleparticleCF}.

\section{Renormalization group equations\label{sec:RG}}

Having calculated the effective action (\ref{Gammafinal}),
we can derive the renormalization group equations. The effective action contains divergent terms in the limit $ma\to 0$. We absorb them by introducing the renormalized coupling constants that are in the following denoted by the subscript $_R$. Characterizing the disorder strength by the dimensionless parameters
\begin{gather}\label{Dcurly}
\mathscr{D}=\pi \rho_2^2 a^3D_b/v^2,\\
\label{Dfcurly}
\mathscr{D}_f=a K^2D_f /v^2,
\end{gather}
we define the renormalized coupling constants as
\begin{gather}\label{1}
\mathscr{D}=Z_b \mathscr{D}_R,\quad \delta=Z(3/2+\delta_R)-3/2,\\
\label{2}
\mathscr{D}_f=Z_f \mathscr{D}_{fR},\quad v=Z v_R,\\
\label{3}
K=Z(3/2+\delta_R),\quad m=m_R.
\end{gather}
Here $Z_b,Z$, and $Z_f$ are the functions that depend on renormalized parameters and on $\lambda$. Those functions are determined in such a way that the effective action expressed in terms of renormalized parameters is finite order by order in an expansion at small $\delta_R$ and $\mathscr{D}_R$, in the limit $a\to 0$. We find
\begin{align}\label{Zb}
Z_b=&1-\frac{1}{2}\left[\left(39-54\ln 2+9c_1\right)\mathscr{D}_R+2\delta_R\right]\lambda\notag\\
&+\frac{1}{4}\left(9\mathscr{D}_R+2\delta_R^2\right)\lambda^2 -\mathscr{D}_R(6c_1+c_2)+\mathcal{O}(\mathscr{D}_R\delta_R),\\
\label{Z}
Z=&1-\left[3\mathscr{D}_R+\left(6\ln 2-4\right)\mathscr{D}_R\delta_R\right]\lambda +\frac{3}{2}\mathscr{D}_R\delta_R\lambda^2\notag\\
&+3c_1\mathscr{D}_R+2c_1\mathscr{D}_R \delta_R+\mathcal{O}(\mathscr{D}_R\delta_R^2),\\
\label{Zf}
Z_f=&1-36\pi \mathscr{D}_R^2/\mathscr{D}_{fR}+\mathcal{O}(\mathscr{D}_R^2\delta_R/\mathscr{D}_{fR}).
\end{align}

The renormalization group equations are obtained by requiring that the derivatives of the bare coupling constants with respect to the scale $\ell=-\ln m$ nullify. By differentiating the expressions (\ref{1}) we obtain a system of two coupled linear equations that can be solved in terms of $\dif\mathscr{D}_R/\dif \ell$ and $\dif\mathscr{\delta}_{R}/\dif \ell$, yielding
\begin{align}\label{D}
&\frac{\dif\mathscr{D}_R}{\dif\ell}=-2\mathscr{D}_R\delta_R+A\mathscr{D}_R^2 +\mathcal{O}(\mathscr{D}_R^2\delta_R),\\
\label{delta}
&\frac{\dif\delta_R}{\dif\ell}=-9\mathscr{D}_R+B\mathscr{D}_R\delta_R +\mathcal{O}(\mathscr{D}_R^2).
\end{align}
The constants  $A=54\ln 2-39-9c_1$ and $B=6-18\ln2+9c_1$ are nonuniversal. They depend on $c_1$ that is determined by the choice of the infrared cutoff function [see Eq.~(\ref{S0-rep}) for one possibility]. Since this choice is not unique, the form of the correlation function (\ref{G}) beyond distances of the order of $m^{-1}$ is nonuniversal. Hence the constant $c_1$ in Eq.~(\ref{a1}) is also nonuniversal, see Appendix \ref{appendixa1}. However, the sum
\begin{align}
\label{invariant}
A+B=36\ln2-33
\end{align}
is a universal number that characterizes our model. We discuss below its role.

It is interesting to note that similar equations were derived for the action $S_0+S_b$ [see Eqs.~(\ref{S0-rep-full}) and (\ref{Sb-rep})] in a different context, namely how a higher dimension affects the transition.\cite{herbut} This work employs $1 + \epsilon$ expansion. The dimension $\epsilon$ plays the role of a regulator for the theory. However, in that case a different universal combination of the parameters $A+B$ appears.

In renormalization group equations (\ref{D}) and (\ref{delta}) we have determined the first subleading terms, being proportional to $A$ and $B$. The term $\mathcal{O}(\mathscr{D}_R^2)$ in Eq.~(\ref{delta}) is controlled by the last term in Eq.~(\ref{eq:divGamma2}) that is beyond two-loop. Setting $A=B=0$ in Eqs.~(\ref{D}) and (\ref{delta}) and using $\delta_R=K_R-3/2$, we obtain the lowest order scaling equations that correspond to those first obtained by Giamarchi and Schulz. \cite{Giamarchi+87EPL,Giamarchi+88} We note that those references use the parameter $1/K$ instead of our $K$.

The other two renormalization group equations are
\begin{gather}
\label{Df}
\frac{\dif\mathscr{D}_{fR}}{\dif\ell}=0+\mathcal{O} (\mathscr{D}_R^2\delta_R),\\\label{tilt} \frac{\dif}{\dif\ell}\left(\frac{v_R}{K_R}\right)=0.
\end{gather}
The last equation turns out to be valid beyond our perturbative calculation. The absence of renormalization to the parameter $v/K$ in our model (\ref{Hamiltonian}) is an exact result, due to the statistical symmetry $\varphi(x,\tau)\to\varphi(x,\tau)+w(x)$, where $w(x)$ is an arbitrary function, of the disordered part of the action.\cite{Schulz+88,Hwa+94} Therefore, Eq.~(\ref{tilt}) is an exact result valid at all orders. The renormalization of forward scattering part of disorder has no consequences for the localization properties of the system.\footnote{From $Z_f$ one can obtain the equation ${\dif\mathscr{D}_{fR}}/{\dif\ell}=-144\pi\mathscr{D}_R^2\delta_R
$. It determines very small depletion of the bare value $\mathscr{D}_{fR}$ that is proportional to $\delta_R^5$ at the transition. We neglect this effect.}

Let us analyze the renormalization group equations. We see that the possible fixed points $(\mathscr{D}_R^*,\delta_R^*)$ in Eqs.~(\ref{D}) and (\ref{delta}) are
(i) $ (0,0)$, (ii) $(0, \delta_R^*>0)$, and (iii) $(18/AB,9/B)$.  The fixed points (iii) are not in the domain of applicability of our calculation as we now show. We assumed that both $\mathscr{D}_R$ and $\delta_R$ are much smaller than unity and thus $9/|B|\ll1$. Using the condition (\ref{invariant}), we obtain $\mathscr{D}_R^*=18/AB\approx -18/B^2<0$. Since $\mathscr{D}_R^*$ measures the disorder strength that must be non-negative, see Eq.~(\ref{definitionD}), the fixed point (iii) is unphysical.

Now we consider the fixed points (i) and (ii). The solution of the flow equations (\ref{D}) and (\ref{delta}) up to third order in $\delta_R$ can be expressed as
\begin{align}\label{solution}
\delta_R^2-9\mathscr{D}_R-A\delta_R\mathscr{D}_R+2(A+B)\delta_R^3/27=C,
\end{align}
where $C$ is an arbitrary constant. The case $C<0$ corresponds to the insulating phase and $C>0$ to the superfluid phase.
Thus, the fixed point (i) marks the transition between the superfluid and the insulating phase. The superfluid is characterized by a line of fixed points (ii) where renormalized disorder strength is zero and $K_R>3/2$. The insulating phase cannot be described using the perturbative calculation since the strength $\mathscr{D}_R$ of disorder grows with increasing the length scale, becoming higher than unity at a finite scale, when our perturbation theory becomes unapplicable.

Using Eq.~(\ref{solution}) we obtain the connection between the parameters $\mathcal{D}_R$ and $\delta_R$ at the critical line determined by the condition $C=0$:
\begin{align}\label{criticallineD}
\mathscr{D}_R=(\delta_R/3)^2+ (2B-A)\delta_R^3/243+\mathcal{O}(\delta_R^4).
\end{align}
Substituting this expression in Eq.~(\ref{delta}), at large scales we obtain
\begin{align}\label{flowdeltal}
&\delta_R(\ell)= \frac{1}{\ell}+\frac{(A+B)}{27}\frac{\ln\ell}{\ell^2}+\mathcal{O}\left(\frac{1}{\ell^2}\right).
\end{align}
At the critical line, the flow of $\delta_R(\ell)$ is universal at large scales at two leading orders, see Eq.~(\ref{flowdeltal}). The term of the order of $1/\ell^2$ depends on the bare parameter $\delta$. Despite the scaling equations (\ref{D}) and (\ref{delta}) contain nonuniversal parameters $A$ and $B$, their universal combination (\ref{invariant}) determines $\delta_R(\ell)$. This has important consequences for the universality of correlation functions, as we discuss below. The flow of the disorder strength is nonuniversal beyond the lowest order, see Eq.~(\ref{criticallineD}).

The correlation length close to the critical line from the insulating side ($C<0$) takes the form $\xi\propto \exp(\pi/\sqrt{9\mathscr{D}_R-\delta_R^2})[1+\mathcal{O}(\delta_R)]$. Therefore, higher-order corrections do not affect $\xi$ in an essential way and the transition is of Berezinskii-Kosterlitz-Thouless type. \cite{Giamarchi+88,Giamarchi+87EPL} In the superfluid phase, the correlation length is infinite.

\section{Generator of connected correlations\label{sec:generator}}

In this section we calculate another useful functional that is called the generator of connected correlations, which will be used in the following sections to calculate the correlation functions. Let us consider an action
\begin{align}\label{Ssss}
S(\varphi)=S_0(\varphi)+gV(\varphi),
\end{align}
where the action $S$ is expressed as a sum of the quadratic part $S_0$ and the anharmonic part $V$. For the action (\ref{Ssss}), the generator of connected correlations $W(J)$ is defined by the relation \cite{Zinn-Justin}
\begin{align}\label{W}
{e}^{{W}(J)}=\int\Dif\varphi {e}^{-S(\varphi)/\hbar+J\varphi},
\end{align}
where $J$ denotes an arbitrary source field. The expression (\ref{W}) can be transformed by introducing a new field $\chi$, such that $\varphi=\chi+GJ$. We then obtain $S_0(\varphi)-\hbar\varphi J=S_0(\chi)-S_0(GJ)$, where $G$ is the propagator such that we have $S_0(\varphi)/\hbar=\langle\varphi|G^{-1}|\varphi\rangle$/2 in the shorthand notation. Equation (\ref{W}) thus becomes
\begin{align}\label{Gamma???}
{e}^{W(J)}=&\int\Dif\chi \exp\left[-\frac{S_0(\chi)}{\hbar}+\frac{S_0(GJ)}{\hbar}-g \frac{V(GJ+\chi)}{\hbar}\right]\notag\\
=&Z_0\exp\left[\frac{S_0(GJ)}{\hbar}\right] \left\langle\exp\left[-g\frac{V(GJ+\chi)}{\hbar}\right] \right\rangle_{\chi},
\end{align}
where $\langle\ldots\rangle_\chi$ denotes an average with respect to the quadratic action $S_0(\chi)$. At small $g$ we can use the cumulant expansion, yielding to
\begin{align}\label{Wgeneral}
W(J)=&\ln Z_0+\frac{S_0(GJ)}{\hbar} -\frac{g}{\hbar}\left\langle V(GJ+\chi)\right\rangle_{\chi} +\mathcal{O}(g^2).
\end{align}
Another expression for $W(J)$ that directly follows from the definition (\ref{W}) is
\begin{align}\label{Wconnection}
W(J)=\ln Z+\ln\left\langle \exp(J\varphi)\right\rangle_{S},
\end{align}
where $Z$ is the partition function of the model and $\langle \ldots\rangle_S$ denotes the average with respect to the action $S$. Combining the last two expressions for $W(J)$ we obtain
$\left\langle \exp(J\varphi)\right\rangle_S$ expressed in powers of the anharmonic coupling $g$.

Now we use the expressions (\ref{Wgeneral}) and (\ref{Wconnection}) for our model (\ref{Srep}). We should have in mind that the above derivation is written in shorthand notation where we suppressed all internal indices (e.g.,~replica indices) and $GJ$ is determined by the structure of the theory. For our model (\ref{Srep}), the abbreviation $GJ$ denotes
\begin{align}
\sum_\beta\int\dif x'\dif \tau' G_{\alpha\beta}(x-x',\tau-\tau')J_\beta(x',\tau'),
\end{align}
where $G_{\alpha\beta}(x,\tau)$ is defined in Eq.~(\ref{Galphabeta}).
The generator of connected correlations then has the form
\begin{align}\label{Wcumulant}
W(J)=\ln Z_0+W_0+W_1+\mathcal{O}(D_b^2)
\end{align}
where
\begin{align}\label{W0}
W_0(J)=&\frac{1}{\hbar}\left[S_0(GJ)+S_f(GJ)\right],\\
\label{W1}
W_1(J)=&-\frac{1}{\hbar}\left\langle S_b(GJ+\chi)\right\rangle_{\chi}.
\end{align}
The final expression for the correlation function with an arbitrary source field $J_\alpha(x,\tau)$ is
\begin{align}\label{correlation:final}
&\left\langle\!\left\langle {e}^{\sum_\alpha\int\dif x\dif\tau J_\alpha(x,\tau) \varphi_\alpha(x,\tau)}\right\rangle\!\right\rangle=\notag\\ &\qquad\qquad \frac{Z_0}{Z} \exp{\left[W_0(J)+W_1(J)+\mathcal{O}(D_b^2)\right]},
\end{align}
where the average $\langle\langle\ldots\rangle\rangle$ is with respect to the replicated action $S_{\mathrm{rep}}$, see Eq.~(\ref{Srep}). Note that $Z=Z_0$ in the replica limit $n\to0$, where $Z_0=\int \prod_{\alpha=1}^n\mathcal{D}\varphi _{\alpha} \exp{\left(-S_0/\hbar-S_f/\hbar\right)}$ and $Z=\int \prod_{\alpha=1}^n\mathcal{D}\varphi_{\alpha} \exp{\left(-S_{\mathrm{rep}}/\hbar\right)}$. In the following sections we consider special choices of the source field in Eq.~(\ref{correlation:final}) to evaluate the density-density and the single-particle correlation functions.

Comparing Eqs.~(\ref{W1}) and (\ref{Gamma1def}) we note similarities between the perturbative expansions of $W$ and the effective action $\Gamma$.\cite{ledoussal+13ristivojevic} Either directly calculating or using the this correspondence, we obtain
\begin{align}\label{W0-calculated}
W_0(J)=&\frac{1}{2}\sum_{\alpha\beta}\int\dif x\dif\tau\dif x'\dif\tau' J_\alpha(x,\tau)\notag\\
&\times G_{\alpha\beta}(x-x',\tau-\tau')J_\beta(x',\tau'),\\ \label{W1-calculated}
W_1(J)=&\rho_2^2 D_b e^{-4G(0,0)}\sum_{\alpha\beta}\int\dif x\dif\tau\dif\tau' \cos(2\Theta_{\alpha\beta})\notag\\ &\times\bigg\{\left[{e}^{4G(0,\tau-\tau')}-1\right] \delta_{\alpha\beta}+1\bigg\},\\
\Theta_{\alpha\beta}=&\int\dif x''\dif\tau'' \bigl[G(x-x'',\tau-\tau'') J_\alpha(x'',\tau'')\notag\\
&-G(x-x'',\tau'-\tau'')J_\beta(x'',\tau'') \bigr].
\end{align}

\section{Density-density correlation function\label{sec:correlations}}

Let us consider the oscillatory part of density-density correlation function. It can be expressed as $2\rho_2^2\cos(2\pi\rho_0x)R_1(x,\tau)$, where
\begin{align}\label{R1}
R_1(x,\tau)=& \overline{\langle e^{2i(\varphi(x,\tau)-\varphi(0,0))}\rangle}_H.
\end{align}
Here $H$ denotes the Hamiltonian (\ref{Hamiltonian}) and the overbar stands for the disorder average. A naive way to calculate this correlation function is to use the renormalized Hamiltonian that is at large scales determined by the fixed point values of the coupling constants. In the present model, the superfluid phase is defined as a phase where the anharmonic coupling constant $D_b$ renormalizes to zero, while the fixed point value of the Luttinger liquid parameter is $K^*>3/2$. This phase is effectively described by the quadratic theory. Using the renormalized parameters, we obtain for the correlation function in the superfluid phase \begin{align}\label{R1naive}
R_1^{\mathrm{SF}}(x,\tau)=e^{-2\frac{K^2}{v^2}D_f|x|} \left(\frac{a}{\sqrt{x^2+(v^*\tau)^2}}\right)^{2K^*},
\end{align}
where we used that $K/v=K^*/v^*$ is not renormalized due to the exact symmetry of our model, as discussed earlier.
At the transition, the fixed point is given by $D_{b}^*=0$ and $K^*=3/2$, and thus one would infer that the power law exponent in Eq.~(\ref{R1naive}) is $3$. As we show below using a more systematic approach, this value for the exponent is correct, but the naive result (\ref{R1naive}) does not predict all the terms at the transition. We find that the correlation function $R_1$ acquires additional logarithmic corrections.

In the following we consider the static correlation function $R_1(x)=R_1(x,0)$ at the transition. A systematic way to obtain the correlation function along the critical line is to consider its renormalization group flow. This leads us to the Callan-Symanzik equation. Its solution is\cite{Amit}
\begin{align}\label{eq:procedure}
R_{1}(x)={e}^{\int_0^{\ell_x} \dif\ell\gamma_{1}(\ell)} R_{1}(a;\delta_R(\ell_x),\mathscr{D}_R(\ell_x),\ldots),
\end{align}
where $\ell_x=\ln(x/a)\gg 1$ and the ellipsis denotes other coupling constants. The right-hand side of Eq.~(\ref{eq:procedure}) contains two terms. The term $R_{1}(a;\delta_R(\ell_x),\mathscr{D}_R(\ell_x),\ldots)$ corresponds to $R_1(a)$ where the coupling constants of the Hamiltonian are taken at the scale $\ell_x$. This term saturates into a constant at large $x$. The function $\gamma_1$ in the exponent of the first term on the right-hand side of Eq.~(\ref{eq:procedure}) is defined as
\begin{align}\label{gammmm}
\gamma_{1}=\frac{\partial\ln Z_{R_{1}}}{\partial \ln a},
\end{align}
where $Z_{R_{1}}$ is the factor that multiplicatively removes $a$-dependent divergences from $R_{1}(x)$, as we now explain.

In order to find $\gamma_1$, the procedure is as follows. \cite{Amit} (i) We calculate $R_1(x,\tau=0)$ of Eq.~(\ref{R1}) in a perturbation theory at small bare coupling $D_b$. (ii) Once we have the perturbative result for $R_1(x)$ expressed in terms of bare parameters (such as $D_b$ and $\delta$), we reexpress them as functions of the renormalized ones, Eqs.~(\ref{1})-(\ref{3}). (iii) The obtained expression will contain divergent behavior at $a\to0$. Those divergent contributions can then be multiplicatively renormalized by introducing $Z_{R_1}$. (iv) We obtain $\gamma_1$ using Eq.~(\ref{gammmm}) and employ Eq.~(\ref{eq:procedure}) to find the final result.

We begin with the step (i). We use the replica formalism and the connection between the correlation functions
\begin{align}\label{R1 connection}
\overline{\langle e^{2i(\varphi(x,\tau)-\varphi(0,0))}\rangle}_H=\lim_{n\to 0} {\langle\langle e^{2i(\varphi_\gamma(x,\tau)-\varphi_\gamma(0,0))}\rangle\rangle},
\end{align}
where $\gamma$ is an arbitrary replica index and $\langle\langle\cdots\rangle\rangle$ denotes the average with respect to the replicated action $S_{\mathrm{rep}}$, Eq.~(\ref{Srep}). The right-hand side of Eq.~(\ref{R1 connection}) we calculate using  Eq.~(\ref{correlation:final}) and the source field
\begin{align}\label{Jchoice}
J_\alpha(x',\tau')=2\,i \left[\delta(x'-x)- \delta(x')\right]\delta(\tau')\delta_{\alpha\gamma}.
\end{align}
Substituting Eq.~(\ref{Jchoice}) in the expression (\ref{W0-calculated}), we find
\begin{align}
W_0=&4G_{\gamma\gamma}(x,0)-4G_{\gamma\gamma}(0,0) \notag\\
=&-\frac{2 K^2 D_f}{v^2}|x|- \left(\frac{3}{2}+\delta\right)\ln\left(\frac{x^2+a^2}{a^2}\right),
\end{align}
where we employed (\ref{Gab}), (\ref{Gsmall}), and (\ref{G0small}) and used the replica $n\to0$ and $m\to 0$ limits.  Substituting Eq.~(\ref{Jchoice}) in (\ref{W1-calculated}), in the limit $m\to 0$ and $n\to 0$ yields
\begin{align}
\label{W1-calculated-special}
W_1=&\rho_2^2 D_b \int\dif  x'\dif\tau'\dif\tau''\frac{1}{\left[1+ v^2(\tau'-\tau'')^2/a^2\right]^{\frac{3}{2}+\delta}} \Biggl\{-1\notag\\ &+\left[\frac{(x'-x)^2+v^2\tau''^2+a^2} {(x'-x)^2+v^2\tau'^2+a^2}\,\,\frac{x'^2+v^2\tau'^2+a^2} {x'^2+v^2\tau''^2+a^2}\right]^{\frac{3}{2}+\delta} \Biggr\}\notag\\
=&\frac{9}{4}\mathscr{D}\ln^2\left(\frac{x^2}{a^2}\right) +c_4\mathscr{D}\ln\left(\frac{x^2}{a^2}\right) +\mathcal{O}\left(\mathscr{D}\delta\right),
\end{align}
with $c_4$ being a numerical constant unimportant for us, as will become clear below. Here we recall the definition (\ref{Dcurly}) and use $\lim_{n\to 0}\sum_{\alpha}\cos(A\delta_{\alpha\gamma})=\cos A-1$. We have thus obtained the perturbative expansion
\begin{align}\label{R1bare}
R_1(x)=&{e}^{-2{\mathscr{D}_f}|x|/a}\left(\frac{a^2}{x^2+a^2} \right)^{\frac{3}{2}+\delta}\biggl[1 +\frac{9}{4}\mathscr{D}\ln^2\left(\frac{x^2}{a^2} \right)\notag\\
& +c_4\mathscr{D}\ln\left(\frac{x^2}{a^2} \right) +\mathcal{O}(\mathscr{D}\delta)\biggr].
\end{align}
The expression in the square brackets in Eq.~(\ref{R1bare}) is the perturbation theory in the small anharmonic term (\ref{Sb-rep}). Therefore, one would expect that the factor $1$ in the square brackets,
which arises from the harmonic theory, is always larger than the remaining perturbative corrections, but this is not the case at large $x\gg a$. However, this issues is resoled by introducing the renormalized parameters.

In step (ii) we express the bare parameters in terms of the renormalized ones. For the correlation function, the lowest order expressions in $\lambda$ suffice:
\begin{gather} \label{delta 1order} \delta=\delta_R+9(c_1-\lambda)\mathscr{D}_R/2+ \mathcal{O}(\mathscr{D}_R\delta_R),\\
\label{D 1order}
\mathscr{D}=\mathscr{D}_R+\mathcal{O}(\mathscr{D}_R\delta_R),\\
v=v_R\left[1-3\mathscr{D}_R\lambda+3c_1\mathscr{D}_R +\mathcal{O}(\mathscr{D}_R\delta_R)\right],\\
\mathscr{D}_{f}=\mathscr{D}_{fR}+\mathcal{O}(\mathscr{D}_{R}^2).
\end{gather}
Using these results, we find
\begin{align}\label{R1renormalized}
R_1(x)=&{e}^{-{2\mathscr{D}_{fR}}\frac{|x|}{a}}
\left(\frac{x^2}{a^2}\right)^{-\frac{3}{2}-\delta_R} \biggl[1-\frac{9}{4}\mathscr{D}_R\lambda^2 \notag\\ &+\frac{1}{2}\left(9c_1-2c_4\right) \mathscr{D}_R\lambda\biggr] \bigl[1+9\mathscr{D}_R \ln^2(cm x)\notag\\
&+\mathscr{D}_R(2c_4-9c_1)\ln (cm x) +\mathcal{O}(\mathscr{D}_R\delta_R)\bigr].
\end{align}
It is important to note that all $x$-dependent divergencies in the limit $a\to 0$ have canceled after introducing the renormalized parameters.

In step (iii) we introduce the multiplicative factor $Z_{R_1}$ that contains all $a$-dependence in Eq.~(\ref{R1renormalized}), such that $Z_{R_1}R_1(x)$ becomes $a$-independent. It reads
\begin{align}
Z_{R_1}=&e^{2\mathscr{D}_{fR}\frac{|x|}{a}}(ma)^{-3-2\delta_R}  \notag\\ &\times\left[1+\frac{9}{4}\mathscr{D}_R\lambda^2 -\frac{1}{2}\left(9c_1-2c_4\right) \mathscr{D}_R\lambda\right].
\end{align}

In the final step (iv) we determine the anomalous dimension function
\begin{align}\label{eq:gammazz}
\gamma_{1}=\frac{\partial \ln Z_{R_{1}}}{\partial\ln a}=-2\mathscr{D}_{fR}e^\ell -3-2\delta_R(\ell)+\mathcal{O}(\mathscr{D}_R,\delta_R^2),
\end{align}
where $\ell=\ln(x/a)$ denotes the scale at which we consider the system. We point out that the running scale $\ell$ is different from the scale $\ell_x$ introduced just below Eq.~(\ref{eq:procedure}). This is because $x$ in the left hand side of Eq.~(\ref{eq:procedure}) and $x$ in the left hand side of Eq.~(\ref{R1renormalized}) have different meanings. The former denotes the real physical distance introduced by the correlation function (\ref{R1}), while the latter determines the scale $\ell\leq \ell_x$. After combining Eqs.~(\ref{delta}) and (\ref{criticallineD}), we
find
\begin{align}\label{integrategamma}
\int_{0}^{\ell_{x}} \dif\ell \gamma_{1}(\ell)=&-2\mathscr{D}_{fR}(e^{\ell_x}-1) -3\ell_x \notag\\&+ \int_{\delta_R(0)}^{\delta_R(\ell_{x})} \frac{\dif\delta_R}{\delta_R}\frac{2} {1+\mathcal{O}(\delta_R)}\notag\\
=&-2\mathscr{D}_{fR}\frac{|x|}{a} -3\ln{\left(\frac{|x|}{a}\right)} +2\ln{\left(\delta_R\right)}
\Big|_{\delta_R(0)}^{\delta_R(\ell_{x})}.
\end{align}
Using Eq.~(\ref{flowdeltal}) for large scale behavior of $\delta_R$ in the previous expression, Eq.~(\ref{eq:procedure}) enables us to obtain the final result at $|x|\gg a$
\begin{align}\label{R1final}
  R_{1}(x)&\sim {e}^{-{2\mathscr{D}_{fR}}\frac{|x|}{a}} \left(\frac{a}{|x|}\right)^{3}
  \ln^{-2}\left(\frac{|x|}{a}\right)\notag\\ &\!\!\times\left\{1+\frac{2(A+B)}{27}\frac{\ln[\ln ( |x|/a)]}{\ln(|x|/a)}+\mathcal{O}\left(\frac{1}{\ln(|x|/a)}\right)\right\},
\end{align}
where $A+B=36\ln2-33$ [see Eq.~(\ref{invariant})] and $\mathscr{D}_{fR}$ denotes the renormalized dimensionless strength of forward scattering. At weak disorder where our study is applicable, $\mathscr{D}_{fR}\approx \mathscr{D}_{f}$, where the latter parameter is defined in Eq.~(\ref{Dfcurly}).

At criticality, we have found a multiplicative logarithmic correction to the density-density correlation function (\ref{R1naive}) that is obtained long ago. \cite{Giamarchi+88} We point out that logarithmic correction $\ln^{-2}{(x/a)}$ in Eq.~(\ref{R1final}) arises from the lowest order renormalization group equations. Second loop contributions in renormalization group equations give rise to the subleading correction in Eq.~(\ref{R1final}) proportional to $A+B$. Despite the fact that our model has nonuniversal flow equations (\ref{D}) and (\ref{delta}), the correlation function (\ref{R1final}) at criticality has the universal form. In Eq.~(\ref{R1final}), omitted terms in the curly bracket that are of the order $1/\ell_x$ may depend on the bare parameters, such as $\delta$. These terms are produced by the neglected higher order terms $\mathcal{O}(\mathscr{D}_R,\delta_R^2)$ in Eq.~(\ref{eq:gammazz}).

\section{Single particle correlation function\label{sec:singleparticleCF}}

In this section we consider the single-particle correlation function
\begin{align}\label{R2beg}
R_2(x)=\overline{\langle\Psi(x)\Psi^\dagger(0)\rangle}_H.
\end{align}
Using the density-phase representation,\cite{haldane81prl} where $\Psi^\dagger(x)=\sqrt{\rho}\,e^{-i\theta(x)}$, we express it as
\begin{align}\label{R2}
R_2(x)\approx\rho_0\overline{\langle e^{i[\theta(x)-\theta(0)]} \rangle}_H.
\end{align}
In order to calculate the correlation function (\ref{R2}) at the criticality, we employ the replica formalism similarly as in the preceding section, see in particular Eq.~(\ref{R1 connection}). Introducing
$R_2(x)\approx \rho_0 \lim_{n\to 0} r_2(x)$, our starting expression becomes
\begin{align}\label{R2start}
r_2(x)=&\left\langle\! \left\langle {e}^{i\left[\theta_\gamma(x,0)-\theta_\gamma(0,0)\right]} \right\rangle\! \right\rangle\notag\\
=& \frac{1}{Z_f}\int\Dif\varphi_1\ldots \Dif\varphi_n\Dif\theta_1\ldots \Dif\theta_n{e}^{-\frac{S_{\mathrm{rep}}}{\hbar}}\notag\\ &\times{e}^{\sum_\alpha\int\dif x'\dif\tau' \theta_\alpha(x',\tau')j_\alpha(x',\tau')},
\end{align}
where
$j_\alpha(x',\tau')=i\left[\delta(x'-x)-\delta(x')\right] \delta(\tau')\delta_{\alpha\gamma}$ and $S_{\mathrm{rep}}$ is given by Eq.~(\ref{Srep}). Here $\gamma$ denotes an arbitrary replica index. Unlike in the previous considerations, here we need the quadratic part of the action that depends both $\theta$ and $\varphi$ fields. Therefore in the replicated action (\ref{Srep}) we use $S_0$ given by the expression (\ref{S0-rep-full}). In Eq.~(\ref{R2start}), $Z_f$ is the abbreviation for the partition function. The evaluation of $r_2(x)$ is conveniently done in Fourier space. We use
\begin{align}
j_\alpha(k,\omega)=i\left({e}^{-i k x}-1\right)\delta_{\alpha\gamma},
\end{align}
and integrate-out $\theta_\alpha$ fields from the numerator and the denominator in the expression (\ref{R2start}). We then obtain
\begin{widetext}
\begin{align}\label{R2start1}
r_2(x)=& \frac{1}{Z}\int\Dif\varphi_1\ldots \Dif\varphi_n\exp{\left[-\frac{1}{2}\sum_{\alpha\beta}\int\frac{\dif k\dif\omega}{(2\pi)^2}\frac{\varphi_\alpha(k,\omega) \varphi_\beta(-k,-\omega)} {G_{\alpha\beta}(k,\omega)}\right]}\notag\\ &\times\exp{\left\{\sum_\alpha\int\frac{\dif k\dif\omega}{(2\pi)^2}\left[\frac{\pi}{2vK} \frac{j_\alpha(k,\omega)j_\alpha(-k,-\omega)}{k^2}  -\frac{i}{vK}\frac{\omega}{k}\varphi_\alpha(k,\omega) j_\alpha(-k,-\omega)\right] \right\}} {e}^{-\frac{S^{}_b}{\hbar}}
\end{align}
\end{widetext}
where we used Eq.~(\ref{S0SF}), while $G_{\alpha\beta}(k,\omega)$ is defined in Eq.~(\ref{Galphabeta}). Using the notation of Sec.~\ref{sec:generator}, the previous expression can be rewritten as
\begin{align}\label{R2und}
&r_2(x)=\left\langle\!\left\langle {e}^{\sum_\alpha\int\dif x\dif \tau \varphi_\alpha(x,\tau) J^\theta_\alpha(x,\tau)} \right\rangle\! \right\rangle\notag\\ &\times\exp{\left\{-\sum_\alpha\int\frac{\dif k\dif\omega}{(2\pi)^2}\left[\frac{\pi vK}{2} \frac{J^\theta_\alpha(k,\omega) J^\theta_\alpha(-k,-\omega)}{\omega^2}\right] \right\}},
\end{align}
where the new source field $J^\theta_\alpha(x,\tau)$ is defined via its Fourier transform
\begin{align}\label{Jtheta}
J^\theta_\alpha(k,\omega)=-\frac{i}{vK} \frac{\omega}{k} j_\alpha(k,\omega)=\frac{1}{vK}\frac{\omega}{k} \left({e}^{-i k x}-1\right)\delta_{\alpha\gamma}.
\end{align}

The remaining calculation is conceptually similar to the one from the previous section. The expression (\ref{R2und}) we evaluate using the results of Sec.~\ref{sec:generator}, in particular Eq.~(\ref{correlation:final}). We start with $W_0$. Expressing Eq.~(\ref{W0-calculated}) in Fourier space, we obtain
\begin{align}
W_0(J)=\frac{1}{2}\sum_{\alpha\beta}\int\frac{\dif k\dif\omega}{(2\pi)^2}J_\alpha(k,\omega) G_{\alpha\beta}(k,\omega) J_\beta(-k,-\omega).
\end{align}
For the source field (\ref{Jtheta}) that occurs in Eq.~(\ref{R2und}), we obtain
\begin{align}\label{W0theta}
&W_0(J^\theta_\alpha)-\sum_\alpha\int\frac{\dif k\dif\omega}{(2\pi)^2}\left[\frac{\pi vK}{2} \frac{J^\theta_\alpha(k,\omega) J^\theta_\alpha(-k,-\omega)}{\omega^2}\right] =\notag\\
&-\frac{\pi}{vK}\int\frac{\dif k\dif\omega}{(2\pi)^2}\frac{1-\cos\left(k x\right)}{k^2+\omega^2/v^2}
=-\frac{1}{4K}\ln\left(\frac{x^2}{a^2}\right),
\end{align}
at $x\gg a$. We emphasize that the previous lowest order result is obtained directly in the limit $m\to 0$, using the propagator (\ref{Galphabeta}) at $m=0$. We also note that due to $J^{\theta}_{\alpha}\propto \omega$, the off-diagonal part in replica indices of the propagator (\ref{Galphabeta}) does not appear in Eq.~(\ref{W0theta}). Therefore, the lowest order perturbation theory result (\ref{W0theta}) for $r_2(x)$ does not depend on the strength of the forward scattering part of the disorder potential $D_f$.

In fact, a more general statement is that $D_f$ does not appear in $r_2(x)$ at any order. This could be inferred  using the following argument.\cite{Giamarchi}
We introduce a new field $\widetilde{\varphi}(x)=\varphi(x)-K\int_0^x\eta(y) dy/\hbar v$ in the starting Hamiltonian (\ref{Hamiltonian}). Let us denote by $\widetilde{H}$ our Hamiltonian expressed in terms of the fields $\widetilde\varphi$ and $\theta$ that satisfy the standard bosonic commutation relation. The Hamiltonian $\widetilde{H}$ is, up to a constant, equal to the sum of $H_0$ [Eq.~(\ref{eq:llham})] and $H_d|_{\eta=0}$ [Eq.~(\ref{eq:Hd})], where $\varphi$ is replaced by $\widetilde\varphi$. Since disorder potential in $H_d$ is short-range correlated and since $\xi(x)$ and $\eta(x)$ are independent, see Eq.~(\ref{eq:mixed}), the shift proportional to $\int_0^x\eta(x)\dif x$ introduced in $\widetilde\varphi$ will not appear in the replicated action corresponding to $\widetilde{H}$. This replicated action is a sum $S_0+S_b$, Eqs.~(\ref{S0-rep-full}) and (\ref{Sb-rep}), where $\varphi$ is replaced by $\tilde{\varphi}$. We note that $S_f$ term (\ref{Sf-rep}) is not contained in the replicated action corresponding to $\widetilde{H}$. Thus $R_2(x)$ does not depend on $D_f$ at any order in the the perturbation theory in $D_b$. However, in the evaluation of $R_1(x,\tau)$ from the previous section, see Eq.~(\ref{R1}), one should account for the difference between $\varphi$ and $\widetilde\varphi$, that leads to the exponential term in Eq.~(\ref{R1final}).

Next we calculate $W_1(J^\theta)$ that we need in the expansion of Eq.~(\ref{R2und}). It is given by Eq.~(\ref{W1-calculated}) for the source field (\ref{Jtheta}). In the limits $m\to 0$ and $n\to 0$, the term that has two replica summations in Eq.~(\ref{W1-calculated}) nullifies, while only one term with one replica summation remains. Employing
\begin{align}
\int\dif k\dif\omega \frac{\omega}{k}\frac{{e}^{i (k x+\omega\tau)}}{\omega^2+v^2 k^2} =-2\pi\arctan\frac{x}{v\tau},
\end{align}
in Eq.~(\ref{W1-calculated}), after some algebra one obtains
\begin{align}\label{compl}
W_1(J^\theta)=&\rho_2^2 D_b\int\frac{\dif x'\dif\tau'\dif\tau''} {\left[1+v^2(\tau'-\tau'')^2/a^2\right]^{\frac{3}{2}+\delta}}
\notag\\ &\times \Bigg\{\left[\sqrt{\frac{A_1A_4 }{A_2A_3}}-\frac{v^2(\tau'-\tau'')^2 x^2} {2\sqrt{A_1A_2A_3A_4}}\right]-1\Bigg\},
\end{align}
where
\begin{gather}
A_1=x'^2+v^2\tau'^2,\quad A_3=(x'-x)^2+v^2\tau'^2,\\
A_2=x'^2+v^2\tau''^2,\quad
A_4=(x'-x)^2+v^2\tau''^2.
\end{gather}
After evaluation Eq.~(\ref{compl}) becomes
\begin{align}
W_1(J^\theta)=-\frac{1}{4}\mathscr{D} \ln^2\left(\frac{x^2}{a^2}\right)
+c_{5}\mathscr{D} \ln\left(\frac{x^2}{a^2}\right)
+\mathcal{O}\left(\mathscr{D}\delta\right),
\end{align}
where $c_5$ is some constant.

We have therefore obtained the expression for $r_2(x)$ at the lowest nontrivial order in $\mathscr{D}$,
\begin{align}\label{R2bare}
r_2(x)=&\left(\frac{a}{|x|}\right)^{\frac{1}{3+2\delta}} \biggl[1-\frac{1}{4}\mathscr{D} \ln^2\left(\frac{x^2}{a^2}\right) +c_{5}\mathscr{D}\ln\left(\frac{x^2}{a^2}\right)\notag\\ &+\mathcal{O}\left(\mathscr{D}\delta\right)\biggr].
\end{align}
Substituting the renormalized parameters given by Eqs.~(\ref{delta 1order}) and (\ref{D 1order}) in the previous expression, we can express the obtained result as
\begin{align}\label{R2renormalized}
r_2(x)=&\left(\frac{|x|}{a}\right)^{-\frac{1}{3+2\delta_R}}\left[1+\frac{1}{4}\mathscr{D}_R \lambda^2 -(c_{5}+\frac{c_1}{2}) \mathscr{D}_R\lambda \right]\notag\\ &\times\bigl[1-\mathscr{D}_R\ln^2(cm|x|)\notag\\
&+(2c_{5}+c_1) \mathscr{D}_R\ln(cm|x|) +\mathcal{O}(\mathscr{D}_R\delta_R) \bigr].
\end{align}
We recall $\lambda=\ln(c^2m^2a^2)$.  One should notice that all $x$ dependent divergencies in the limit $a\to 0$ in Eq.~(\ref{R2bare}) have canceled once we expressed the bare coupling constants by the renormalized ones. Namely, the combination of the term proportional to $\ln^2 (x^2/a^2)$ term in the right-hand side of Eq.~(\ref{R2bare}) combines with $-\mathscr{D}_R\lambda\ln(|x|/a)$ that arises after expanding the power $1/(3+2\delta)$ in Eq.~(\ref{R2bare}) reexpressed in terms of the renormalized parameters.

The expression (\ref{R2renormalized}) contains divergence when $a\to 0$ that could be removed multiplicatively by
\begin{align}
Z_{r_2}=(ma)^{-\frac{1}{3+2\delta_R}}\left[1-\frac{1}{4}\mathscr{D}_R\lambda +\frac{1}{2}(2c_{5}+c_1) \mathscr{D}_R \right],
\end{align}
so that the renormalized correlation function $Z_{r_2} r_2(x)$ has no dependence of $a$. The anomalous dimension function then becomes
\begin{align}
\gamma_{2}=\frac{\partial \ln Z_{r_2}}{\partial\ln m}= -\frac{1}{3}+\frac{2}{9}\delta_R(\ell) +\mathcal{O}(\mathscr{D}_R,\delta_R^2),
\end{align}
where we have used the scaling equations (\ref{D}) and (\ref{delta}). We notice that the previous result could be expressed also as $-1/2K_R$, where $K_R=3/2+\delta_R$. After evaluating $\int_0^{\ell_x} \gamma_{2}(\ell)\dif\ell$, using the similar integration as in Eq.~(\ref{integrategamma}), we finally obtain
\begin{align}\label{R2final}
R_{2}(x)\sim& \rho_0\left(\frac{a}{|x|}\right)^{1/3} \ln^{2/9}\left(\frac{|x|}{a}\right)\notag\\
&\times
  \left[1-\frac{2(A+B)}{243} \frac{\ln{[\ln{(|x|/a)}]}}{\ln{(|x|/a)}}\right].
\end{align}

The single-particle correlation function (\ref{R2beg}) at the transition between the superfluid and the disordered phase takes the form (\ref{R2final}). The power law in Eq.~(\ref{R2final}) with the universal exponent $1/3$ was found in Refs.~\onlinecite{Giamarchi+87EPL,Giamarchi+88}. It can be understood simply by using the quadratic theory described by the fixed point value of the Luttinger liquid parameter $K_R^*=3/2$. Here we find an additional logarithmic prefactor with another universal exponent $2/9$. This result follows from the lowest order renormalization group equations, \cite{Giamarchi+87EPL,Giamarchi+88} as we revealed here. The renormalization group equations at two-loop order enabled us to find additional additive logarithmic corrections in Eq.~(\ref{R2final}), proportional to the universal constant of our model $A+B$, see Eq.~(\ref{invariant}).

\section{Bosonic ladder system}\label{sec:ladder}

So far we considered such bosonic system where changes of particles at neighboring sites (in a discrete lattice picture) are constrained to be in absolute value either zero or one, in order to keep the density fluctuations small and be able to use the bosonization procedure. A more realistic physical situation would require to have possibility for larger particle fluctuations. Such a behavior is certainly expected in the weak interaction side of the phase diagram. To implement this idea within the technique of bosonization is a challenging problem. For clean systems some similar situations were already encountered. \cite{berg+08PhysRevB.77.245119} Although a full solution of this question is still beyond reach, a first step in this direction is to consider a bosonic two-leg ladder system where each of the two chains is exposed to the {same} disorder potential. The rung on which charge fluctuations could go from zero to two even for hard core bosons, could then mimic a ``site'' in a single chain for which large charge fluctuations would be allowed. The rung interaction serves as an additional degree of freedom that should mimic a possibility in a single chain to have larger site to site particle number fluctuations. We thus study such a model in this section.

We consider a two-leg ladder system of disordered bosons. Assuming only the rung interaction and the same disorder in each chain, the Hamiltonian in the continuous description is
\begin{align}\label{HL}
H_L=\sum_{j=1}^2 H(\theta_j,\varphi_j)+V\int\dif x \rho_1(x)\rho_2(x).
\end{align}
Here $H$ is the Hamiltonian (\ref{Hamiltonian}), while the density is given by Eq.~(\ref{density}). If the system is at incommensurate  fillings, the terms from the interrung interaction that are important in the long wavelength limit are
\begin{align}
V\int\dif x \left[\frac{\partial_x\varphi_1(x) \partial_x\varphi_2(x)}{\pi^2} +2\rho_2^2\cos(2\varphi_1(x)-2\varphi_2(x)) \right].
\end{align}
The product of gradients is diagonalized by introducing
\begin{align}\label{phic}
\varphi_c=(\varphi_1+\varphi_2)/\sqrt{2},\quad \theta_c=(\theta_1+\theta_2)/\sqrt{2},\\
\label{phis}
\varphi_s=(\varphi_1-\varphi_2)/\sqrt{2},\quad \theta_s=(\theta_1-\theta_2)/\sqrt{2},
\end{align}
leading to
\begin{align}\label{HL1}
H_L=&\frac{\hbar}{2\pi}\int\dif x\left\{v_c K_c[\partial_x{\theta_c}(x)]^2 +\frac{v_c}{K_c}[\partial_x{\varphi_c}(x)]^2 \right\}\notag\\
&+\frac{\hbar}{2\pi}\int\dif x\left\{v_s K_s[\partial_x{\theta_s}(x)]^2 +\frac{v_s}{K_s}[\partial_x{\varphi_s}(x)]^2 \right\}\notag\\
&+2V\rho_2^2\int\dif x\cos(\sqrt{8}\varphi_s) -\frac{\sqrt{2}}{\pi}\int\dif x \eta(x)\partial_x\varphi_c\notag\\ &+2\rho_2\int\dif x \left[\xi^*(x) {e}^{i\sqrt{2}\varphi_c}\cos(\sqrt{2}\varphi_s) +\text{h.c.}\right].
\end{align}
Here
\begin{gather}\label{conn1}
v_c K_c=v_s K_s=vK,\\
\label{conn2}
\frac{v_c}{K_c}=\frac{v}{K}\left(1+\frac{VK}{\pi\hbar v}\right),\quad \frac{v_s}{K_s}=\frac{v}{K}\left(1-\frac{VK}{\pi\hbar v}\right).
\end{gather}
The Hamiltonian (\ref{HL1}) quite resembles to the one of interacting spinful electrons in a disordered potential.\cite{Giamarchi+88} Note that, of course, here there is no reason a priori to be on a special separatrix imposed by SU(2) symmetry. Thus, we could use the renormalization group equation derived in Ref.~\onlinecite{Giamarchi+88},
\begin{gather}\label{eq1}
\frac{\dif K_{cR}}{\dif\ell}=-\frac{1}{2}\frac{K_{cR}^2 v_{cR}}{v_{sR}}\mathcal{D}_R,\quad
\frac{\dif K_{sR}}{\dif\ell}=-\frac{1}{2}K_{sR}^2\left(\mathcal{D}_R+y_R^2 \right),\\
\label{eq2}
\frac{\dif v_{cR}}{\dif\ell}=-\frac{v_{cR}^2K_{cR}} {2v_{sR}}\mathcal{D}_R,\quad
\frac{\dif v_{sR}}{\dif\ell}=-\frac{v_{sR}K_{sR}}{2}\mathcal{D}_R,\\
\label{eqyyyy}
\frac{\dif y_R}{\dif\ell}=(2-2K_{sR})y_R-\mathcal{D}_R,\\
\label{eq3}
\frac{\dif \mathcal{D}_R}{\dif\ell}=(3-K_{cR}-K_{sR}-y_R)\mathcal{D}_R,
\end{gather}
in terms of the dimensionless quantities
\begin{gather}\label{Dy}
\mathcal{D}=\frac{D_b}{\pi^2 \rho_2 v_s^2} \left(\frac{v_s}{v_c}\right)^{K_c},\quad y=\frac{V}{\pi\hbar v_s}.
\end{gather}
The expressions (\ref{conn1}) and (\ref{conn2}) determine the initial values of the parameters $K_c$, $K_s$, $v_c$, and $v_s$ in terms of the initial values of $K$, $v$ and $y$ as
\begin{gather}
K_c=\frac{K\sqrt{2}}{\sqrt{2-K^2y^2+Ky \sqrt{4+K^2y^2}}},\quad
v_c=\frac{vK}{K_c},\\
K_s=\frac{K}{2}\left(Ky+\sqrt{4+K^2y^2}\right),\quad
v_s=\frac{vK}{K_s}.
\end{gather}

\begin{figure}
\includegraphics[width=0.8\columnwidth]{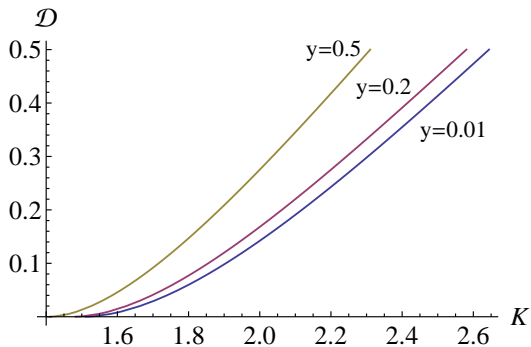}
\caption{\label{fig2}Phase boundary between the Luttinger liquid phase and the disordered phase for the bosonic ladder system (\ref{HL}) as a function of the bare parameters $K$ and $\mathcal{D}$ is given for three initial values of the rung coupling $V$, corresponding to $y=0.01$, $y=0.2$, and $y=0.5$. The region of large $K$ corresponds to the Luttinger liquid phase, where the anharmonic coupling constants $\mathcal{D}$ and $y$ renormalize to zero at large scales.}
\end{figure}

\begin{figure}
\includegraphics[width=0.8\columnwidth]{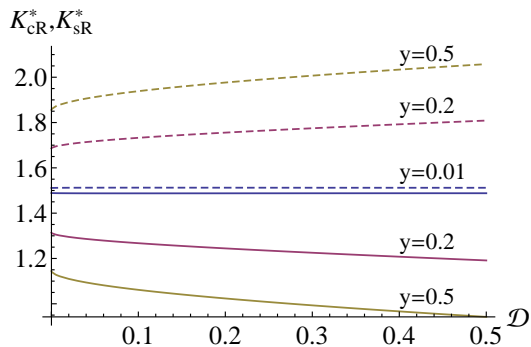}
\caption{\label{fig3}Solid (dashed) lines represent fixed point value $K_{cR}^*$ ($K_{sR}^*$) for the parameter $K_c$ ($K_s$) at the transition as a function of the bare disorder strength $\mathcal{D}$, for different values of the bare coupling $y$. At the transition we have $K_{cR}^*+K_{sR}^*=3$, however, each of the parameters depends on bare disorder and rung couplings, leading to the nonuniversal exponents in various correlation functions.}
\end{figure}

In order to find the renormalized fixed point values of the Luttinger liquid parameters $K_c$ and $K_s$ at the transition,
we numerically solved the system (\ref{eq1})-(\ref{eq3}). Strong quantum fluctuations renormalize the anharmonic coupling constants $y$ and $\mathcal{D}$ to zero. In this limit, the model (\ref{HL}) exhibits a line of Gaussian fixed points. The phase boundary between the latter and the disordered phase is given in Fig.~\ref{fig2} for different values of the rung constant as a function of $K$ and the initial disorder strength. In Fig.~\ref{fig3} we give the values of renormalized values of the Luttinger liquid parameters at the transition. Unlike the fermionic case where the only fixed point with $V_R^*=0$ is the one characterized by $K_{sR}^*=1$ due to the spin symmetry,\cite{Giamarchi+88} in the present case we have a more complicated situation. The absence of such a symmetry leads to higher values for $K_{sR}^*$ at the transition that depend on the bare values of $y$ and $\mathcal{D}$. However, at the transition we now have the universal value for the sum, $K_{cR}^*+K_{sR}^*=3$, while each of the two parameters is nonuniversal. This ultimately leads to the nonuniversal correlation functions at the transition due to nonuniversal exponents, in contrast to the non-ladder case, examined in previous sections. Let us mention that in the special case of zero rung coupling, corresponding to $y=0$, the Hamiltonian (\ref{HL1}) [cf. Eq.~(\ref{HL})] describes two decoupled chains. In this case the scaling equation (\ref{eqyyyy}) does not apply, since the rung coupling that is initially absent cannot be created by the disorder. We discuss this in the Appendix \ref{appendixladder}.

\section{Discussions\label{sec:discussion}}

\begin{figure}
\includegraphics[width=0.8\columnwidth]{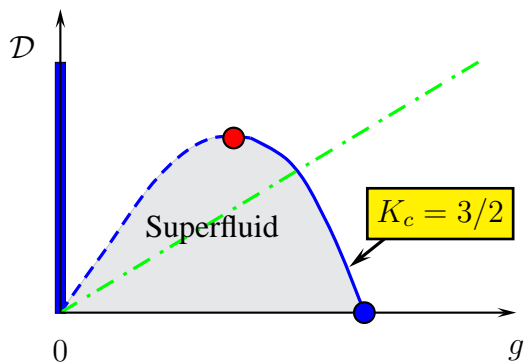}
\caption{\label{fig1} Possible phase diagram of a disordered one-dimensional Bose gas as a function of the boson repulsion $g$ and the disorder $\mathscr{D}$. The green dashed-dotted line schematically indicates the boundary below which a bosonization description of such a problem is guaranteed (namely the disorder is smaller than the chemical potential). In this region
the superfluid/Bose-glass transition, denoted by the solid blue line,
is described by universal exponents. The question of the nature and critical behavior of the transition in the regime for which bosonization cannot be directly applied is yet open (see text). One possibility (not shown on the figure) is that the exponent remain universal along the whole superfluid/Bose-glass line, leading to a single Bose glass phase. The only singular line is then the noninteracting bosons ($g=0$) which are localized by rare events of the disorder and indicated by the blue box.
The other possibility (shown in the figure), is that if at small interaction, there are non-universal exponents along the dashed blue line, then this forces the existence of a critical point on the superfluid/Bose-glass boundary between the non-universal and universal regime.  Consequences for the various phases (superfluid or Bose glass) are still unknown. For example it could lead to the existence of two distinct Bose glass phases.}
\end{figure}

Finally, in this section we discuss the phase diagram. The new terms in the renormalization group flow equations lead to additional insight in the physics of the superfluid/Bose-glass transition for the model (\ref{Hamiltonian}). At the fist order in $D_b$ no off-diagonal (in replica space) renormalization of the quadratic action could be generated, for trivial reasons. The second order is the lowest order at which such terms might in principle appear. From Eq.~(\ref{Gammafinal}) we see that no such terms are generated, except for terms describing forward scattering process, which have no consequences for the localization properties of the system.  The quadratic part of the effective action thus remains, even to the order $D_b^2$ essentially diagonal in replica. The transition between the superfluid and the Bose-glass phases can thus be fully characterized by the two parameters $K$ and $D_f$, and not, like in other models such as the classical Cardy-Ostlund model \cite{Cardy+82}, by a full set of variables (corresponding to the off-diagonal terms) which enter into the correlation functions and can affect the exponents of correlation functions at the transition. This important result, in agreement with the analysis of vortex proliferation of Ref.~\onlinecite{svistunov+96PhysRevB.53.13091},
proven here directly from the renormalization group flow, is a consequence of the time independence of the disorder.\cite{Giamarchi+96PhysRevB.53.15206} Indeed, for two independent replicas $\alpha\neq\beta$ before averaging over disorder one obtains
\begin{align} \label{eq:aver}
\langle \varphi_\alpha(x,\tau) \varphi_\beta(x',\tau') \rangle = &\langle \varphi_\alpha(x,\tau) \rangle \langle \varphi_\beta(x',\tau') \rangle\notag\\
=&\langle \varphi_\alpha(x,0) \rangle \langle \varphi_\beta(x',0) \rangle
\end{align}
since the disorder does not depend on $\tau$. The correlation (\ref{eq:aver}) is thus time independent.
Thus terms such as $\int \dif x\dif x' \dif\tau [\partial_\tau \varphi_\alpha(x,\tau)][\partial_\tau \varphi_\beta(x',\tau)]$ cannot appear in the effective action, at any order, since they would lead to time dependence for Eq.~(\ref{eq:aver}). This has important consequences for the physical properties at the separatrix between the phase for which the disorder is irrelevant ($D_b \to 0$) and the phase for which the disorder is relevant (the Bose glass phase). The absence of such off diagonal replica terms thus leads to a universal value of the parameter $K$, namely $K_c=3/2$, and correlation functions will thus decay with a universal exponent at the transition. Our analysis thus confirms that the superfluid/Bose-glass transition around the value $K=3/2$, i.e., for intermediate interactions, has a generic universal exponent in a finite region around that point. This puts stringent constraints on the phase diagram, as schematized in Fig.~\ref{fig1}.
Indeed, this results which uses the bosonization representation of the disordered Bose gas is guaranteed to work when the disorder is weaker than the chemical potential. This means that for a good part of the superfluid/Bose-glass boundary the exponent will remain constant. What happens for larger disorder or weaker interactions is still an open question. \cite{Svistunov05,Vojta,Pollet13,Pielawa+13,pollet+14PhysRevB.89.054204}
One possibility is of course that the exponent remains universal along the whole line. In that case, there is most likely a unique Bose glass phase. The only singular line of the phase diagram would be in that case, the non-interacting line $g=0$ for which the bosons are localized by rare events of the random potential in a finite region of space. However, if for weak interactions one can obtain non-universal exponents at the transition superfluid/Bose-glass as discussed in Refs.~\onlinecite{Altman08,Altman+10}, it is then impossible to smoothly connect the two separatrices between the superfluid and the Bose glass. It would thus imply that there is on the separatrix a critical point above which the exponents would be universal and below which they would vary with parameters. This would have interesting consequences for the various phases in the system, which are still to be understood. One possible scenario could be two superfluid or two Bose glass phases, as was argued as one of the possible scenarios in Ref.~\onlinecite{Giamarchi+88}.
Finding a good order parameter for such a scenario is, however, not easy. One possibility might be the moments of superfluid stiffness distributions. An interesting question is whether a similar mechanism can occur inside the superfluid as well (see, e.g.,~Ref.~\onlinecite{Gurarie+08PhysRevLett.101.170407}). This is clearly a question which is largely open and will need further analyses both on the analytical and numerical front.

On the experimental front probing the phase diagram is a challenging question. The superfluid/Bose-glass transition can be seen directly from the superfluid correlation function, which will go from a divergent power law behavior
in the superfluid phase, to an exponentially decreasing one in the localized phase. Varying the strength of the disorder and of the interactions
allows to probe the universality of the exponent at the transition. Such behavior can be probed in cold atomic systems or in magnetic insulators. \cite{giamarchi+08naturephysics, Zheludev+10PhysRevB.81.060410, Yamada+11PhysRevB.83.020409,Zapf+12nature} Cold atoms offer the advantage of the control over disorder and interactions. However, optical lattice systems suffer from the inhomogeneity due to the confining potential, which strongly complicates the analysis of the exponent. Atom chips realizations are up to now limited to relatively small interactions ($K\sim 40$). In all these systems the finite size of the system is also an additional complication. Magnetic insulators are very homogeneous and allow precise control and measurement of the boson density and compressibility. Controlling the disorder is, however, more challenging. No doubt that for both types of system these difficulties will be overcome in the future allowing reliable answer on the above points.

\section{Conclusions}\label{sec:conclusions}

In conclusion, we have computed to two-loop order the renormalization group equations for the system of interacting disordered bosons described by the model (\ref{Hamiltonian}).  We have shown, both from the renormalization group and from general arguments that in the regime described by bosonization (intermediate interactions) the separatrix between the superfluid and the Bose glass phase is characterized by a universal exponent. We computed the single particle and density-density correlations at the transition. We found logarithmic corrections in those quantities. Our calculation uses bosonization and therefore also applies to disordered fermions.

We also considered a disordered bosonic ladder system, with correlated disorder along the rungs described by the model (\ref{HL}). The presence of two chains exposed to the same disorder enabled us to mimic larger local fluctuations of the boson occupations on a single site than in the bosonization description (\ref{Hamiltonian}). The rung coupling serves as an additional degree of freedom with respect to a single bosonic chain. In contrast to a single chain,
we find that the rung coupling leads to the nonuniversal exponent at the superfluid/Bose-glass transition. This goes well with the idea that increasing the disorder with respect to interactions, which naturally allows for larger density fluctuations, could indeed lead to non-universal exponents. However, the full solution of a model with only a single bosonic species still remains to be made.

\acknowledgements

We acknowledge discussions with E. Altman, G. Refael, B. Svistunov, and T. Vojta. This work was  supported by PALM Labex and by the Swiss NSF under MaNEP and Division II.

\appendix

\section{Evaluation of $a_1$\label{appendixa1}}

Here we give the details about the evaluation of $a_1$ defined by Eq.~{(\ref{a1def})}. In order to obtain the renormalization group equations, we need to find all the divergent contributions
in the limit $a\to 0$ in the integral on the right-hand side of Eq.~{(\ref{a1def})}. They originate from the region of integration where $\tau$ is around zero. Therefore, $a_1$ can be written as
\begin{align}
a_1=\frac{2}{v^3}\int_0^{\Delta}dy y^2\left[e^{4G(0,y/v)}-1\right]+\mathrm{f.t.}
\end{align}
where
\begin{align}\label{acondition}
a\ll \Delta\ll (cm)^{-1},
\end{align}
while $\mathrm{f.t.}$ stands for the remaining terms that are finite in the limit $a\to 0$. Using the small-$\tau$
expansion of $G(0,y/v)$ given by Eq.~(\ref{Gsmall}), we obtain \begin{align}\label{evaluationa1}
a_1=&\frac{2}{(cmv)^3}\int_0^{\Delta} \frac{\dif y\, y^2}{(y^2+a^2)^{3/2}}\{ 1-\delta \ln[c^2m^2(y^2+a^2)]\notag\\&+\mathcal{O}(\delta^2)\}+\mathrm{f.t.}\notag\\
=&\frac{-2\lambda+ \delta\lambda^2-4(\ln2-1)\delta\lambda+\mathcal{O}(\delta^2) }{2(cmv)^3}+\mathrm{f.t.},
\end{align}
where $\lambda=\ln (c^2m^2a^2)$, $c={e}^{\gamma_E}/2$, with  $\gamma_E$ being the Euler constant. In Eq.~(\ref{evaluationa1}) we have evaluated the divergent contributions to order $\delta$. This is because we want to obtain the renormalization group equations at two-loop order. However, at two-loop order we also need to find the finite term at $a\to 0$ and at order $\delta^0$, which is the constant $c_1$ appearing in Eq.~(\ref{a1}). Thus considering the defining equation for $c_1$,
\begin{align}
\int\dif\tau \tau^2 \left[e^{4G(0,\tau)}-1\right]=\frac{-\lambda+c_1}{(cmv)^3}
\end{align}
in the limit $a\to 0$ and $\delta\to 0$, we obtain
\begin{align}
c_1=&\lim_{a\to 0}\left\{\lambda+(c m v)^3\int \dif\tau \tau^2\left[{e}^{3K_0(m\sqrt{v^2\tau^2+a^2})} -1\right]\right\}\notag\\ \approx&\, 7.17.
\end{align}
The constant $c_1$ depends on the behavior of the correlation function (\ref{G}) at distances beyond $m^{-1}$. Since the  behavior in this region is determined by the cutoff function, $c_1$ is nonuniversal. However, the scaling equations (\ref{D}) and (\ref{delta}) at the criticality lead to the universal behavior.

\section{Evaluation of $b_1$\label{appendixb1}}

In this appendix we evaluate $b_1$ defined by Eq.~(\ref{defb1}). We want to find the renormalization group equation for the strength of the backward scattering part of the disorder up to its quadratic dependence, see Eq.~(\ref{D}). Since  $b_1$ in $\Gamma_2$ is multiplied by $D_b^2$ [see Eq.~(\ref{eq:divGamma2})] and since close to the transition $D_b\propto\mathscr{D}_R\sim \delta_R^2\approx\delta^2$ [see Eqs.(\ref{Dcurly}) and (\ref{solution})] at two-loop order is sufficient to evaluate $b_1$ at $\delta=0$, which is used in the following.

Introducing ${\bf{z}}=(0,v\tau')$ and ${\bf{r}}=(x,v\tau)$, the integral (\ref{defb1}) becomes
\begin{align}\label{eq:b1}
b_1=&\frac{1}{v^2}\int_{-\infty}^{\infty}d z\int_{0}^{\infty}dr\,r \int_0^{2\pi} d\theta\Bigl\{e^{4[G({\bf{z}})+ G({\bf{r}})-G({\bf{z}}+{\bf{r}})]}\notag\\&-e^{4G(\bf{z})}
-e^{4G(\bf{r})}-e^{-4G(\bf{z}+\bf{r})}+2\Bigr\},
\end{align}
where $z=v\tau'$, while $\theta$ denotes the angle between $\bf{z}$ and $\bf{y}$. In Eq.~(\ref{eq:b1}) we have used the fact that the term $4[e^{4G(0,\tau-\tau')}][G(x,\tau')-G(x,\tau)]$ from Eq.~(\ref{f2fin}) nullifies after the integration in Eq.~(\ref{defb1}).

In order to evaluate Eq.~(\ref{eq:b1}) and find the divergent terms in $b_1$ the limit $a\to 0$, we split the domain of integration into four regions:
\begin{align*}
&\text{(i)}\;|z|,r<\Delta,\quad &&\text{(ii)}\;|z|<\Delta \; \text{and}\; r>\Delta,\\
&\text{(iii)}\;|z|>\Delta \; \text{and}\; r<\Delta,\quad &&\text{(iv)}\;|z|>\Delta \; \text{and}\; r>\Delta.
\end{align*}
Here we choose $\Delta$ to satisfy the condition (\ref{acondition}). We express $b_1=b_{11}+b_{12}+b_{13}+b_{14}$, where the contribution $b_{1j}$ arises from the region (j).

We start the evaluation considering the region (i). There we can use the small distance expansion (\ref{Gsmall}) of all the propagators appearing in $b_{11}$. Keeping only the terms that could lead to divergent behavior at $a\to 0$, after simple change of coordinates, we obtain
\begin{align}
  b_{11}=&\frac{2}{v^2(cm)^3}\int_0^{\Delta/a} \dif \widetilde{z}\int_1^{\sqrt{(\Delta/a)^2+1}} \dif y\,y  \int_{0}^{2\pi}\dif\theta \biggl\{-\frac{1}{y^3}\notag\\&+ \frac{[\widetilde{z}^2+y^2+2\widetilde{z} \sqrt{y^2-1}\cos{(\theta)}]^{3/2}}{(\widetilde{z}^2+1)^{3/2}y^{3}} -\frac{1}{(\widetilde{z}^2+1)^{3/2}}\biggr\}\notag\\
&+\text{f.t.}
\end{align}
We begin to calculate this integral by expanding $\sqrt{y^2-1}=y+\cdots$. We then perform the integration over $y$, yielding the expression of the form
\begin{align}
b_{11}= &\frac{2}{v^2(cm)^3}\int_0^{w} \dif \widetilde{z}\int_0^{2\pi}\dif \theta \bigl[F(\widetilde{z},\sqrt{w^2+1},\theta)\notag\\
&-F(\widetilde{z},1,\theta)\bigr],\quad w=\Delta/a.
\end{align}
The integral of $F(\widetilde{z},\sqrt{w^2+1},\theta)$ in the large-$w$ limit we calculate by first expanding the function at large $w$, and then performing the two integrations. We find
\begin{align}
b_{11}'=\frac{\pi}{2v^2(c m)^3}\left[18\ln^2 w+(36\ln 2-39)\ln w\right]+\text{f.t.}
\end{align}
The integral of $F(\widetilde{z},1,\theta)$ in the large-$w$ limit we calculate in four steps. First we perform the derivative with respect to $w$ and then expand the obtained expression at large $w$. In the next step we integrate the obtained expression over $\theta$. Finally, we perform an indefinite integration over $w$, yielding
\begin{align}
b_{11}''=\frac{3\pi}{2v^2(c m)^3}\left[3\ln^2 w-(6\ln 2+1)\ln w\right]+\text{f.t.}
\end{align}
Using $b_{11}=b_{11}'-b_{11}''$, we obtain
\begin{align}
b_{11}=\frac{9\pi}{2v^2(cm)^3}\biggl[ \ln^2{\frac{\Delta}{a}}+2(3\ln2-2)\ln{\frac{\Delta}{a}}\biggr]+\text{f.t.}
\end{align}
This result contains all the divergent terms in the limit $\Delta/a\to \infty$.

In the region (ii) we can use the small distance expansion [given by Eq.~(\ref{Gsmall})] only for $G(\bf{z})$, while $G({\bf{z}}+{\bf{r}})$ we  expand as
\begin{align}
G({\bf{z}}+{\bf{r}})=&G({\bf{r}})+G'({\bf{r}}) \frac{h}{2\sqrt{r^2+a^2}}\notag\\
&+\left[\frac{G''({\bf{r}})}{r^2+a^2}-\frac{G'({\bf{r}})}{(r^2+a^2)^{3/2}}
\right]\frac{h^2}{8}+\mathcal{O}(h^3),
\end{align}
where $h=z^2+2zr\cos{\theta}$. After performing the integration over $\theta$ in Eq.~(\ref{eq:b1}), we obtain
\begin{align}
b_{12}=&-\frac{4\pi}{v^2(cm)^3}\int_0^{\Delta}dz\int_{\Delta}^{+\infty}\dif r\, r \frac{z^2}{(z^2+a^2)^{3/2}}\notag\\ &\times\left\{ \frac{G'({\bf{r}})}{r}+G''({\bf{r}})-[G'({\bf{r}})]^2+\mathcal{O}(z^2)\right\} +\mathrm{f.t.}
\end{align}
This leads to
\begin{align}
  b_{12}=&\frac{\pi}{v^2(cm)^3}\left(1-\ln\frac{2\Delta}{a}\right) \left[\frac{15}{2}+9\ln{(cm\Delta)}\right] +\mathrm{f.t.},
\end{align}
where $c=e^{\gamma_E}/2$.

The regions $\text{(iii)}$ and $\text{(iv)}$ do not give divergent contributions. Therefore, $b_1=b_{11}+b_{12}$. Keeping only the divergent terms in the sum gives rise to
\begin{align}
b_1=&\frac{\pi}{2v^2(cm)^3}\biggl[9\ln^2\frac{\Delta}{a}-18 \ln(cm\Delta)\ln\frac{\Delta}{a}\notag\\
&+(54\ln2-51)\ln\frac{\Delta}{a}\biggr]+\text{f.t.}\notag\\
=&\frac{\pi}{2v^2(cm)^3}\biggl[9\ln^2(cma)+(51-54\ln2)\ln(cma)\biggr]+\text{f.t.}
\end{align}
We notice that all the divergent terms proportional to $\Delta$ in the final sum cancel. The latter expression is given in the main text in Eq.~(\ref{b1}), where $c_2$ stands for finite terms in the limit $a\to0$, while we also added a term $\mathcal{O}(\delta)$ to emphasize the accuracy of $b_1$.

\section{Evaluation of $b_2$}\label{appendixb2}

Here we evaluate $b_2$ defined by Eq.~(\ref{defb2}). We are interested in finding the renormalization group equation for the strength of the forward scattering part of disorder up to quadratic dependence in backward scattering strength $D_b$ [see Eq.~(\ref{Df})]. Since $b_2$ appears in $\Gamma_2$  multiplied by $D_b^2$ and since close to the transition $D_b\propto\mathscr{D}_R\sim \delta^2$, see Eqs.~(\ref{eq:divGamma2}) and (\ref{solution}), we calculate the divergent terms in $b_2$  at the order $\delta^0$. Introducing $y=v\tau$ and $z=v\tau'$, the coefficient $b_2$ can be expressed as
\begin{align}\label{eq:b2}
b_2=\frac{1}{v^2}\int dx dy dz x^2 \left[ e^{4G(x,y/v)}-1\right]\left[ e^{4G(x,z/v)}-1\right].
\end{align}
In order to calculate the divergent contribution in the limit $a\to 0$, we can split the region of integration into eight regions determined by $|x|\gtrless \Delta$, $|y|\gtrless \Delta$ and $|z|\gtrless \Delta$, where $\Delta$ satisfies the condition (\ref{acondition}). However, only in the region $|x|,|y|,|z|<\Delta$ the result has a divergent behavior. In this region we can use small distance expansion of the two propagators $G$ appearing in Eq.~(\ref{eq:b2}), given by Eq.~(\ref{Gsmall}). We then obtain
\begin{align}
b_2=&\frac{1}{v^2(cm)^{6}a}\int_{-\Delta/a}^{+\Delta/a}dx dy dz \biggl[\frac{x^2}{(x^2+y^2+1)^{3/2}}
\notag\\&\times\frac{1}{(x^2+z^2+1)^{3/2}}\biggr] +\mathrm{f.t.}+\mathcal{O}(\delta)\notag\\
=&\frac{1}{a}\frac{2\pi}{v^2(cm)^6}+\mathrm{f.t.}+\mathcal{O}(\delta).
\end{align}
Here we have found the divergent terms at the order $\delta^0$, and $\mathrm{f.t.}$ stands for finite terms in the limit $a\to 0$.

\section{The derivation of the renormalization group equations for the ladder system\label{appendixladder}}

In this appendix we give the details about the derivation of the renormalization group equations at the lowest one-loop order for the ladder system described by the Hamiltonian (\ref{HL1}). After performing the disorder average using the replica method, we obtain the action
\begin{align}\label{SL}
\frac{S_L}{\hbar}=\frac{S_0[\varphi_{c}]}{\hbar}+ \frac{S_0[\varphi_{s}]}{\hbar}+ \frac{S_V}{\hbar} + \frac{S_B}{\hbar},
\end{align}
where the quadratic part  $S_0$ for the charge and spin degrees of freedom [see Eqs.~(\ref{phic}) and (\ref{phis})] is given by Eq.~(\ref{S0-rep}). We use the subscripts $_c$ and $_s$ to distinguish the velocities, the Luttinger parameters, and the corresponding fields. The anharmonic terms in Eq.~(\ref{SL}) are
\begin{align}\label{SV}
\frac{S_V}{\hbar}=&\frac{2\rho_2^2V}{\hbar}\sum_\alpha\int \dif x\dif\tau \cos[\sqrt{8}\varphi_{s\alpha}(x,\tau)],\\
\label{SB}
 \frac{S_B}{\hbar}=& -4\rho_2^2 D_b\sum_{\alpha\beta} \int\dif x\dif \tau\dif\tau' \cos[\sqrt{2}\varphi_{c\alpha}(x,\tau) \notag\\ & - \sqrt{2}\varphi_{c\beta}(x,\tau')] \cos[\sqrt{2}\varphi_{s\alpha}(x,\tau)]  \cos[\sqrt{2}\varphi_{s\beta}(x,\tau')].
\end{align}
For simplicity in the following we omit the forward scattering term. It plays no role at the lowest one-loop order, as it can be gauged away using a transformation similar to the one explained in the paragraph below Eq.~(\ref{W0theta}).\cite{Giamarchi+88}

Our aim is to calculate the effective action $\Gamma$ corresponding to the action (\ref{SL}). We use the same procedure as in the main text [Eq.~(\ref{Gamma pert})]. In the present case, the difference is that we have two anharmonic terms given by Eqs.~(\ref{SV}) and (\ref{SB}). At one-loop order we need to evaluate $\Gamma$ at the first two orders in $V$, but only at the lowest order in $D_b$.

The contribution at linear order in $V$ is
\begin{align}
\Gamma_1^{(V)}=2\mathscr{V}\sum_\alpha \int\dif x\dif\tau\cos(\sqrt{8}\varphi_{s\alpha}),
\end{align}
where $\mathscr{V}=(\rho_2^2 V/\hbar)e^{-4G_s(0,0)}$.
Here and in the following we use the notation $G_{s\alpha\beta}(x,\tau)=\langle \varphi_{s\alpha}(x,\tau) \varphi_{s\beta}(0,0)\rangle=G_s(x,\tau)\delta_{\alpha\beta}$, and similarly for $\varphi_c$ fields. We notice that the off-diagonal part of the correlation function is zero, $G_0\equiv 0$, since we omitted the forward scattering term [cf.~Eqs.~(\ref{Gabdef}) and (\ref{Gab})].

At quadratic order, we obtain
\begin{align}\label{Gamma2V}
\Gamma_2^{(V)}=&-\mathscr{V}^2\sum_{\alpha}\sum_{j=\pm 1} \int\dif x\dif\tau \dif x'\dif\tau'\!\left[e^{8j G_s(x-x',\tau-\tau')}-1\right]\notag\\ &\times\cos[\sqrt{8}\varphi_{s\alpha}(x,\tau)-j\sqrt{8}\varphi_{s\alpha}(x',\tau')].
\end{align}
The most relevant contribution from the latter term is
\begin{align}
\Gamma_2^{(V)}=&4\mathscr{V}^2\sum_{\alpha} \int\dif x\dif\tau \left[ \widetilde{a}_1 (\partial_x\varphi_{s\alpha})^2 + \widetilde{a}_2 (\partial_\tau \varphi_{s\alpha})^2\right]+\ldots
\end{align}
where
\begin{align}
\widetilde{a}_1=\int\dif x\dif\tau x^2\left[e^{8G_s(x,\tau)}-1 \right],\\
\widetilde{a}_2=\int\dif x\dif\tau \tau^2 \left[e^{8G_s(x,\tau)}-1 \right].
\end{align}
At one-loop order, we do not need the behavior of $G_s$ at large distances, and thus it is sufficient to use [see Eq.~(\ref{Gsmall})]
\begin{align}
G_s(x,\tau)=-\frac{K_s}{4}\ln[c^2m^2(x^2+v_s^2\tau^2+a^2)],
\end{align}
provided we use the upper boundary of integration of the order of $1/m$. We now introduce the deviation around  $K_s=1$ as
\begin{align}\label{deltasdef}
\delta_s=K_s-1.
\end{align}
At one-loop order, it is sufficient to use $\delta_s=0$ in the integrals. Thus, in the leading order
\begin{align}\label{a1a2tilde}
\widetilde{a}_1=-\frac{\pi\lambda}{2(cm)^4 v_s},\quad \widetilde{a}_2=-\frac{\pi \lambda}{2(cm)^4 v_s^3}.
\end{align}
Here we recall $\lambda=\ln (c^2m^2a^2)$. We emphasize that at one-loop order we do not need neither the constant term nor the terms linear in $\delta_s$ in Eq.~(\ref{a1a2tilde}). Also, we notice that in principle the first term in brackets in Eq.~(\ref{Gamma2V}) should contain the term $-8jG_s(x-x',\tau-\tau')$ that arises from the last term in Eq.~(\ref{Gamma2def}). This term makes the expression to be one-particle irreducible.\cite{Zinn-Justin} However, in practice we do not need $-8jG_s(x-x',\tau-\tau')$ term at one-loop order as it does not affect the divergent parts of $\widetilde{a}_1$ and $\widetilde{a}_2$ given by Eq.~(\ref{a1a2tilde}).

\begin{widetext}
The lowest order term in $D_b$ is
\begin{align}
\Gamma_1^{(B)}=&-\mathcal{B} \sum_{\alpha\beta} \int\dif x\dif \tau\dif\tau' \cos[\sqrt{2}\varphi_{c\alpha}(x,\tau)- \sqrt{2}\varphi_{c\beta}(x,\tau')] \cos[\sqrt{2}\varphi_{s\alpha}(x,\tau)]  \cos[\sqrt{2}\varphi_{s\beta}(x,\tau')]\notag\\
&-\frac{\mathcal{B}}{2}\sum_\alpha\sum_{j=\pm 1} \int\dif x\dif \tau\dif\tau' \cos[\sqrt{2}\varphi_{c\alpha}(x,\tau)- \sqrt{2}\varphi_{c\alpha}(x,\tau')] \cos[\sqrt{2}\varphi_{s\alpha}(x,\tau)+j\sqrt{2}\varphi_{s\alpha}(x,\tau')]\notag\\ &\times\left[e^{2G_c(0,\tau-\tau')-2jG_s(0,\tau-\tau')}-1\right],
\end{align}
where $\mathcal{B}=4\rho_2^2 D_b e^{-2G_c(0,0)-2G_s(0,0)}$. After performing the gradient expansion, the most relevant contributions are
\begin{align}\label{Gamma1B}
\Gamma_1^{(B)}=&-\mathcal{B} \sum_{\alpha\beta} \int\dif x\dif \tau\dif\tau' \cos[\sqrt{2}\varphi_{c\alpha}(x,\tau)- \sqrt{2}\varphi_{c\beta}(x,\tau')] \cos[\sqrt{2}\varphi_{s\alpha}(x,\tau)]  \cos[\sqrt{2}\varphi_{s\beta}(x,\tau')]\notag\\
&-\frac{\mathcal{B}}{2} \widetilde{a}_3 \sum_\alpha\int\dif x\dif \tau \cos(\sqrt{8}\varphi_{s\alpha}) + \frac{\mathcal{B}}{2} \widetilde{a}_4 \sum_\alpha\int\dif x\dif \tau \left[ (\partial_\tau\varphi_{c\alpha})^2+(\partial_\tau\varphi_{s\alpha})^2 \right]+\ldots
\end{align}
\end{widetext}
Here
\begin{gather}\label{a3tilde}
\widetilde{a}_3=\int\dif\tau \left[e^{2G_c(0,\tau)-2G_s(0,\tau)}-1 \right],\\
\label{a4tilde}
\widetilde{a}_4=\int\dif\tau \tau^2 \left[e^{2G_c(0,\tau)+2G_s(0,\tau)}-1 \right].
\end{gather}
In the following we use
\begin{align}
G_c(x,\tau)=-\frac{K_c}{4}\ln[c^2m^2(x^2+v_c^2\tau^2+a^2)],
\end{align}
[see Eq.~(\ref{Gsmall})] and the deviation around $K_c=2$ introduced as
\begin{align}\label{deltacdef}
\delta_c=K_c-2.
\end{align}
We then easily obtain in the leading order
\begin{align}
\widetilde{a}_3=-\frac{v_s\lambda}{c m v_c^2},\quad \widetilde{a}_4=-\frac{\lambda}{(c m)^3 v_c^2 v_s},
\end{align}
where we safely set $\delta_c=0$ at one-loop order.

Finally, we need the term produced by the product of $S_B$ and $S_V$ in $\Gamma_2$. The formal expression is obtained from Eq.~(\ref{Gamma2def}) and takes the form
\begin{align}
\Gamma_2^{(VB)}=&-\frac{1}{\hbar^2}\langle S_V(\varphi+\chi) S_B(\varphi+\chi)\rangle_\chi+\Gamma_1^{(V)}\Gamma_1^{(B)}.
\end{align}
Here we do not account for the last term in Eq.~(\ref{Gamma2def}) that makes the expression to be one-particle irreducible, which is not important for our purpose. After a straightforward evaluation we obtain
\begin{widetext}
\begin{align}
\Gamma_2^{(VB)}=&\frac{\mathscr{V}\mathcal{B}}{2} \sum_{\alpha}\int \dif x\dif\tau\dif\tau' \dif x''\dif \tau'' \cos[\sqrt{2}\varphi_{c\alpha}(x,\tau)-\sqrt{2}\varphi_{c\alpha}(x,\tau')]\Biggl\{ 8\cosh[4G_s(x'',\tau'')]-8\notag\\
&+e^{2G_c(0,\tau-\tau')-2G_s(0,\tau-\tau')}\biggl\{ -2\cos[\sqrt{2}\varphi_{s\alpha}(x,\tau) +\sqrt{2}\varphi_{s\alpha}(x,\tau')]\cos[\sqrt{8}\varphi_{s\alpha}(x'',\tau'')] \notag\\ &+\sum_{j=\pm 1} e^{-4j G_s(x-x'',\tau-\tau'')-4j G_s(x-x'',\tau'-\tau'')}\cos[\sqrt{2}\varphi_{s\alpha}(x,\tau) +\sqrt{2}\varphi_{s\alpha}(x,\tau') +j\sqrt{8}\varphi_{s\alpha}(x'',\tau'')]\biggr\}\notag\\
&+2e^{2G_c(0,\tau-\tau')+2G_s(0,\tau-\tau')} \Bigl\{ -\cos[\sqrt{2}\varphi_{s\alpha}(x,\tau) -\sqrt{2}\varphi_{s\alpha}(x,\tau')]\cos[\sqrt{8}\varphi_{s\alpha}(x'',\tau'')]\notag\\ &+ e^{-4 G_s(x-x'',\tau-\tau'')+4 G_s(x-x'',\tau'-\tau'')}\cos[\sqrt{2}\varphi_{s\alpha}(x,\tau) -\sqrt{2}\varphi_{s\alpha}(x,\tau') +\sqrt{8}\varphi_{s\alpha}(x'',\tau'')]\Bigr\}
\Biggr\}\notag\\
&+2 \mathscr{V}\mathcal{B} \sum_{j=\pm 1}\sum_{\alpha\beta} \int \dif x\dif\tau\dif\tau' \dif x''\dif \tau'' \cos[\sqrt{2}\varphi_{c\alpha}(x,\tau)-\sqrt{2}\varphi_{c\beta}(x,\tau')]\notag\\ &\times\cos[\sqrt{2}\varphi_{s\beta}(x,\tau')]\cos[\sqrt{2}\varphi_{s\alpha} (x,\tau)+j \sqrt{8}\varphi_{s\alpha}(x'',\tau'')]\left[e^{-4jG_s(x-x'',\tau-\tau'')}-1\right].
\end{align}
After the gradient expansion, we find that the one-replica part does not renormalize $\cos(\sqrt{8}\varphi_{s\alpha})$ operator, but renormalizes the parameters in the quadratic part of the action. However, this change is of higher order and thus neglected here. The only contribution arises from the two-replica term at $j=-1$,
\begin{align}
\Gamma_2^{(VB)}=2 \mathscr{V}\mathcal{B} \widetilde{a}_5\sum_{\alpha\beta} \int \dif x\dif\tau\dif\tau' \cos[\sqrt{2}\varphi_{c\alpha}(x,\tau)-\sqrt{2}\varphi_{c\beta}(x,\tau')] \cos[\sqrt{2}\varphi_{s\beta}(x,\tau')]\cos[\sqrt{2}\varphi_{s\alpha} (x,\tau)]+\ldots,
\end{align}
where
\begin{align}
\widetilde{a}_5=\int\dif x\dif\tau \left[e^{4G_s(x,\tau)}-1\right].
\end{align}
The leading order term is
\begin{align}
\widetilde{a}_5=-\frac{\pi\lambda}{(cm)^2 v_s}.
\end{align}

Collecting the above terms, for the model (\ref{SL}) we obtain the effective action
\begin{align}\label{Gammafinalladder}
\Gamma=&\sum_{\alpha} \int\dif x\dif\tau\left\{\frac{v_c}{2\pi K_c}\left[(\partial_x\varphi_{c\alpha})^2+m^2(\varphi_{c\alpha})^2\right]+ \left[\frac{1}{2\pi  K_c v_c} +\frac{\mathcal{B}}{2}\widetilde{a}_4\right] (\partial_\tau\varphi_{c\alpha})^2\right\}\notag\\
&+\sum_{\alpha} \int\dif x\dif\tau\left\{\left[\frac{v_s}{2\pi K_s}+4\mathscr{V}^2\widetilde{a}_1\right] (\partial_x\varphi_{s\alpha})^2+\frac{v_s}{2\pi K_s} m^2(\varphi_{s\alpha})^2+ \left[\frac{1}{2\pi  K_s v_s} +4\mathscr{V}^2\widetilde{a}_2+\frac{\mathcal{B}}{2}\widetilde{a}_4\right] (\partial_\tau\varphi_{s\alpha})^2 \right\}\notag\\
&-\left(\mathcal{B}-2\mathcal{B}\mathscr{V}\widetilde{a}_5\right) \sum_{\alpha\beta} \int\dif x\dif\tau\dif\tau' \cos[\sqrt{2}\varphi_{c\alpha}(x,\tau)-\sqrt{2}\varphi_{c\beta}(x,\tau')] \cos[\sqrt{2}\varphi_{s\alpha}(x,\tau)] \cos[\sqrt{2}\varphi_{s\beta}(x,\tau')]\notag\\
&+\left(2\mathscr{V}-\frac{\mathcal{B}}{2}\widetilde{a}_3\right) \sum_\alpha\int\dif x\dif\tau \cos(\sqrt{8}\varphi_{s\alpha}).
\end{align}
\end{widetext}
The effective action (\ref{Gammafinalladder}) contains divergent contributions in the limit $ma\to 0$. They can be absorbed by defining the renormalized parameters, denoted by the superscript $_R$, introduced by the relations
\begin{align}
&K_c=Z_0 K_{cR},\quad &&\delta_c=Z_0(2+\delta_{cR})-2,\\
&K_s=Z_1 K_{sR},\quad &&\delta_s=Z_1(1+\delta_{sR})-1,\\
&v_c=Z_0 v_{cR},\quad &&v_s=Z_2 v_{sR},\\
&\mathcal{D}=Z_B\mathcal{D}_R,\quad &&y=Z_V y_R,
\end{align}
where the dimensionless parameters measuring the disorder strength $\mathcal{D}$ and the rung coupling $y$ are defined as
\begin{align}\label{Dymy}
  \mathcal{D}=\frac{8\pi \rho_2^2 a^3 D_b}{v_c^2},\quad y= \frac{4\pi\rho_2^2a^2 V}{\hbar v_s}.
\end{align}
The $Z$-factors depend on the renormalized quantities, while the only dependence on $m$ is through $\lambda$. These factors are found by requiring that the effective action expressed in terms of the renormalized quantities does not have divergent terms. At the lowest order we obtain
\begin{align}
&Z_0=1-\frac{v_{cR}}{2v_{sR}}\mathcal{D}_R\lambda,\quad Z_1=1-\frac{1}{4}(y_R^2+\mathcal{D}_R)\lambda,\\
&Z_2=1-\frac{1}{4}\mathcal{D}_R\lambda,\quad
Z_B=1-\frac{1}{2}\left(\delta_{cR}+\delta_{sR}+y_R\right)\lambda,\\
&Z_V=1-\left(\frac{\mathcal{D_R}}{2y_R}+\delta_{sR}\right)\lambda.
\end{align}
The renormalization group scaling equations are now easily obtained by the requirement that the derivative of the bare couplings with respect to the scale $\ell=-\ln m$ nullify. Recalling Eqs.~(\ref{deltacdef}) and (\ref{deltasdef}), it yields
\begin{gather}\label{eqm1}
\frac{\dif \delta_{cR}}{\dif\ell}=-\frac{2 v_{cR}}{v_{sR}}\mathcal{D}_R,\quad
\frac{\dif \delta_{sR}}{\dif\ell}= -\frac{1}{2}\left(y_R^2+\mathcal{D}_R\right),\\
\frac{\dif v_{cR}}{\dif\ell}=-\frac{v_{cR}^2} {v_{sR}}\mathcal{D}_R,\quad
\frac{\dif v_{sR}}{\dif\ell}=-\frac{v_{sR}}{2}\mathcal{D}_R,\\
\frac{\dif y_R}{\dif\ell}=(2-2K_{sR})y_R-\mathcal{D}_R,\\
\label{eqm3}
\frac{\dif \mathcal{D}_R}{\dif\ell}=(3-K_{cR}-K_{sR}-y_R)\mathcal{D}_R,
\end{gather}
where
\begin{gather}
K_{cR}=2+\delta_{cR},\quad K_{sR}=1+\delta_{sR}.
\end{gather}
Equations (\ref{eqm1})-(\ref{eqm3}) are in agreement with the ones derived by a different method in Ref.~\onlinecite{Giamarchi+88} and given by Eqs.~(\ref{eq1})-(\ref{eq3}), provided one expands  the right-hand side of Eqs.~(\ref{eq1}) and (\ref{eq2}) at the lowest order. This is achieved by setting $K_c=2$ and $K_c=1$. We also notice that Ref.~\onlinecite{Giamarchi+88} uses $\rho_2=1/(2\pi a)$. In this case, the definitions in Eqs.~ (\ref{Dymy}) and (\ref{Dy}) coincide.

In the special case of no rung coupling ($V=0)$, the model (\ref{SL}) describes two decoupled bosonic systems exposed to the same disorder. Thus, it is important to verify that the effective action (\ref{Gammafinalladder}) for the model (\ref{SL}) in this special case coincides with the one-loop effective action of the model given by Eqs.~(\ref{S0-rep}) and (\ref{Sb-rep}). In the present case, we study two decoupled system, each of them described by the latter model. This is directly seen from Eq.~(\ref{HL}) at $V=0$. However, the present calculation uses different degrees of freedom, Eqs.~(\ref{phic}) and (\ref{phis}). Since $\mathscr{V}=0$, the only nontrivial part of the effective action that is nonzero is $\Gamma_1^{(B)}$, see Eq.~(\ref{Gamma1B}). Also, $V=0$ implies $K_c=K_s$ and $v_c=v_s$, see Eqs.~(\ref{conn1}) and (\ref{conn2}). Therefore, $\widetilde{a}_3=0$, as now becomes obvious from Eq.~(\ref{a3tilde}). Thus, unlike in the case $V\neq 0$ where the disorder affects the rung coupling, at $V=0$ the disorder can not generate the rung coupling, which is expected. This is not possible to observe from the scaling equations of Ref.~\onlinecite{Giamarchi+88}, or Eqs.~(\ref{eqm1})-(\ref{eqm3}), since they are obtained by accounting for the deviations around $K_c=2$ and $K_s=1$, which is the asymmetry between charge and spin degrees of freedom introduced in the renormalization procedure. At $V=0$, we must treat them symmetrically accounting for the deviation around $K_c=K_s=K=3/2$. However, this does not change the value of $\widetilde{a}_4$ [see Eq.~(\ref{a4tilde})], which is controlled by $K_c+K_s$ that has the same value in both cases $V=0$ and $V\neq 0$.

The effective action in the case $V=0$ has the form (\ref{Gammafinalladder}), provided one sets there $\mathscr{V}=0$ and $\widetilde{a}_3=0$. The parameters $K_c$ and $K_s$ are renormalized in the same way. Also $v_c$ and $v_s$. The divergent contributions are absorbed by the renormalized parameters
\begin{align}
&K_c=K_s=\widetilde{Z}_0 K_{R},\quad &&\delta_c=\delta_s=\widetilde{Z}_0\left(\frac{3}{2}+\delta_{R}\right)-\frac{3}{2},\\
&v_c=v_s=\widetilde{Z}_0 v_{R},\quad &&\mathcal{D}=\widetilde{Z}_B\mathcal{D}_R.
\end{align}
At the lowest order we then find
\begin{align}
&\widetilde{Z}_0=1-\frac{3}{8}\mathcal{D}_R\lambda,\quad &\widetilde{Z}_B=1-\delta_{R}\lambda.
\end{align}
It leads to the renormalization group equations
\begin{gather}\label{eqmsss1}
\frac{\dif \delta_{R}}{\dif\ell}=-\frac{9}{8}\mathcal{D}_R,\quad
\frac{\dif}{\dif\ell}\left(\frac{v_R}{K_R}\right)=0,\\
\label{eqmsss3}
\frac{\dif \mathcal{D}_R}{\dif\ell}=(3-2K_{R})\mathcal{D}_R,
\end{gather}
where
\begin{gather}
K_{R}=\frac{3}{2}+\delta_{R}.
\end{gather}
Those equations are equivalent at one-loop order to the ones derived earlier [Eqs.~(\ref{D}) and (\ref{delta})] if one accounts for the difference in the definition of the dimensionless disorder strength for the two cases in Eqs.~(\ref{Dcurly}) and (\ref{Dymy}).

%

\end{document}